\documentclass[ aps,  twocolumn, nofootinbib, showkeys, notitlepage, superscriptaddress,noeprint]{revtex4-2}

\usepackage{natbib}
\usepackage{graphicx}
\usepackage{dcolumn}
\usepackage{bm}
\usepackage{amsmath}
\usepackage{float}
\usepackage{multirow}
\usepackage{slashed}
\usepackage{xcolor}
\usepackage{physics}
\usepackage{multirow}
\usepackage{gensymb}
\usepackage[colorlinks=true, pdfstartview=FitV, bookmarks=true, bookmarksnumbered=true, breaklinks]{hyperref}
\usepackage{mathtools,braket}

\usepackage{lipsum}  
\usepackage{color}
\usepackage{soul}
\definecolor{blue}{rgb}{0.0, 0.0, 1.0}
\definecolor{red}{rgb}{1.0, 0.0, 0.0}
\definecolor{royalblue}{rgb}{0.0, 0.14, 0.4}
\hypersetup{linkcolor=red, citecolor=blue, urlcolor=blue}

\def\orcid#1{\kern .08em\href{https://orcid.org/#1}{\includegraphics[keepaspectratio,width=0.7em]{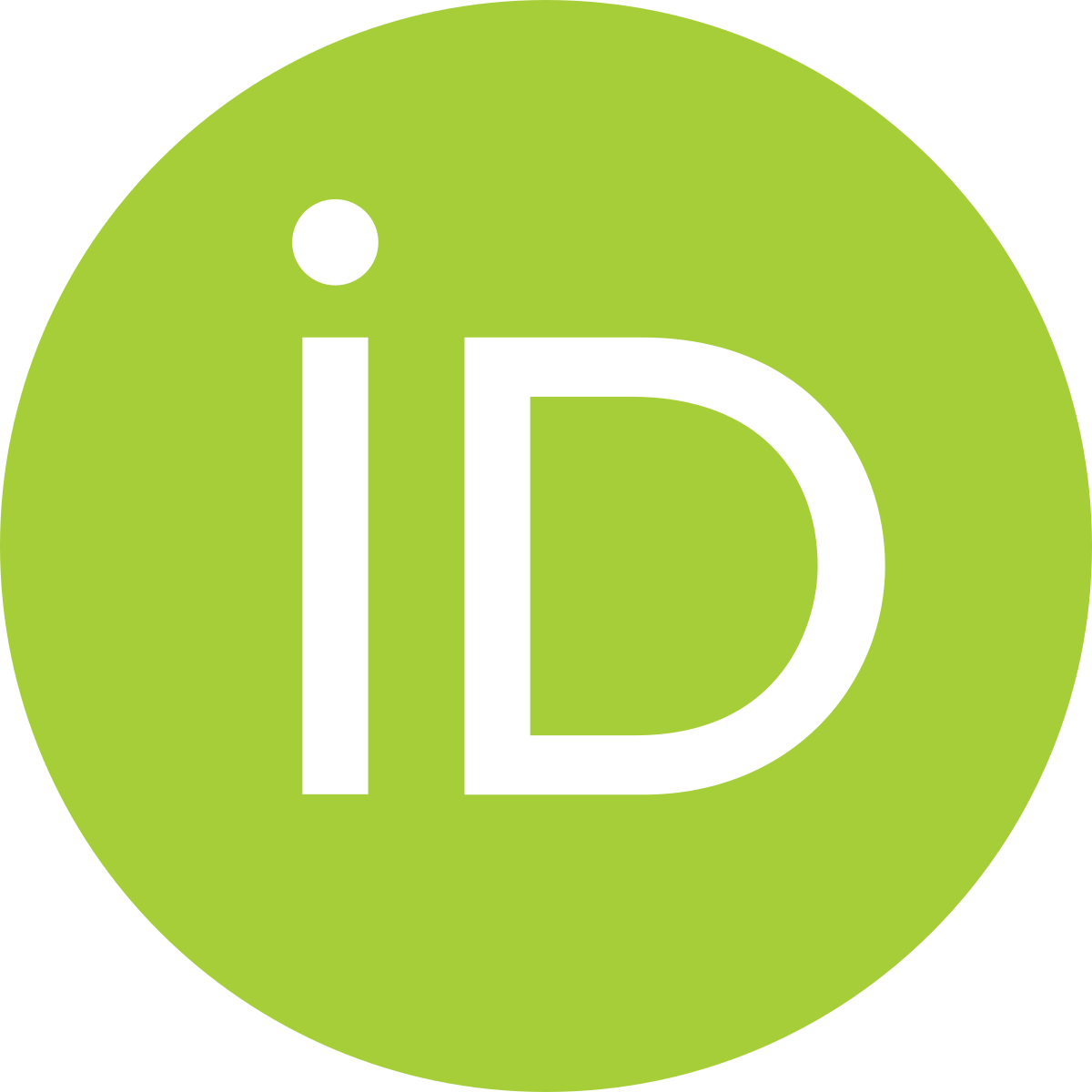}}}

\def\ep{\epsilon}

\def\la{\langle}
\def\ra{\rangle}

\def\be{\begin{equation}}
\def\ee{\end{equation}}
\def\bea{\begin{eqnarray}}
\def\eea{\end{eqnarray}}

\begin{document}
\title{ 
Beyond leading twist: $\rho$ meson decay constants and distribution amplitudes \\ 
in a self-consistent light-front quark model }

\author{Ahmad Jafar Arifi\orcid{0000-0002-9530-8993}}
\email{arifi.jafar@jaea.go.jp}
\affiliation{Advanced Science Research Center, Japan Atomic Energy Agency,  Tokai, Ibaraki 319-1195, Japan}
\affiliation{Research Center for Nuclear Physics, The University of Osaka, Ibaraki, Osaka 567-0047, Japan}

\author{Ho-Meoyng Choi\orcid{0000-0003-1604-7279}}
\email{homyoung@knu.ac.kr}
\affiliation{Department of Physics Education, Teachers College, Kyungpook National University, Daegu 41566, Korea}

\author{Chueng-Ryong Ji\orcid{0000-0002-3024-5186}}
\email{ji@ncsu.edu}
\affiliation{Department of Physics, North Carolina State University, Raleigh, NC 27695-8202, USA}
\date{\today}

\begin{abstract}

In this study, we present a comprehensive analysis of decay constants and chiral-even and chiral-odd distribution amplitudes (DAs) 
up to twist 4 
for the $\rho$ meson in the standard light-front quark model (LFQM) based on the Bakamjian-Thomas (BT) construction. 
For the $\rho$ meson, which possesses both longitudinal $(h=0)$ and transverse $(h=\pm 1)$ polarizations,
two types of decay constants, $f_{\rho}^{\parallel}$ and $f_{\rho}^{\perp}$, arises accordingly.
We demonstrate that these decay constants can be self-consistently extracted from both local ($z^\mu=0$) and nonlocal ($z^\mu\neq 0$) matrix elements $\la 0|{\bar q}(z){\Gamma} q(-z)|\rho(P,h)\ra$, with $\Gamma=(\gamma^\mu, \sigma^{\mu\nu},\gamma^\mu\gamma_5,\bm{1})$, in a manner independent of current components, polarizations, and reference frames.
In particular, we emphasize the role of nonlocal matrix elements involving axial-vector and scalar currents, where mixing between  
$f_\rho^{\parallel}$ and $f_\rho^{\perp}$ occurs. We show that this mixing is consistently resolved through the 
BT construction, ensuring the proper extraction of these decay constants.
Additionally, we investigate the structure of chiral-even DAs 
($\phi_{2;\mathrm{V}}^\parallel, \phi_{3;\mathrm{V}}^\parallel, \psi_{3;\mathrm{A}}^\perp, \phi_{4;\mathrm{V}}^\parallel$) 
and chiral-odd DAs  
($\phi_{2;\mathrm{T}}^\perp, \phi_{3;\mathrm{T}}^\perp, \psi_{3;\mathrm{S}}^\parallel, \phi_{4;\mathrm{T}}^\perp$) 
beyond leading twists, and present  their corresponding $\xi$-moments and Gegenbauer moments.
These results provide deeper insight into the nonperturbative structure of vector mesons and demonstrate the robustness and self-consistency of the LFQM based on the BT framework.

\end{abstract}

\maketitle

\section{Introduction}

Meson distribution amplitudes (DAs) play a fundamental role in hadron physics, encapsulating the internal quark-gluon structure of mesons
and enhancing the computational precision of high-energy processes, form factors, and exclusive decays 
within the collinear factorization 
and perturbative Quantum Chromodynamics (QCD) frameworks~\cite{Lepage:1980fj,ER80,Chernyak:1983ej}. 

As essential non-perturbative inputs, DAs provide critical insights into QCD dynamics and meson properties.
Leading-twist DAs describe the momentum distribution of valence quarks, offering the primary approximation for parton distributions at high momentum.
In contrast, higher-twist DAs incorporate quark-gluon interactions, transverse parton motion, and higher Fock state contributions,
refining our understanding of hadronic processes across different energy scales.

The increasing experimental precision at facilities such as KEKII~\cite{Belle-II:2010dht}, LHC~\cite{LHCb:2012myk}, JLAB~\cite{Dudek:2012vr}, and the future Electron-Ion-Collider (EIC)~\cite{Accardi:2012qut, AbdulKhalek:2021gbh} 
presents new opportunities to explore higher-twist effects in hadronic structures~\cite{Braun:2022gzl,PK18,BBNS,Beppu:2012vi}.
Various theoretical approaches---including QCD sum rules (SRs)~\cite{Ball:1998sk,Ball:1998ff,Ball:2006wn,Ball:2007zt,HWZ04,MPS10,Hu:2024tmc}, light-cone sum rules~\cite{Agaev05}, the instanton vacuum model~\cite{PPRWG,NK06,Liu:2023fpj}, chiral-quark and Nambu--Jona-Lasinio models~\cite{AB02,PR01}, 
the Dyson-Schwinger equation (DSE) approach~\cite{CRS13,SCCRSZ}, and the light-front quark model (LFQM)~\cite{Hwang10,CJ07,C13,C15,Choi:2017uos}---have been employed to study leading- and higher-twist DAs.

In light-front dynamics (LFD)~\cite{BPP}, meson DAs are defined as matrix elements of quark-antiquark bilocal operators on the light cone, 
providing a powerful tool for studying quark-gluon dynamics within hadrons. LFD simplifies the vacuum structure, 
ensures boost invariance, and facilitates factorization theorems, making it an effective framework for exploring high-energy processes and parton distributions.
The LFQM, based on LFD, treats mesons as bound states of a constituent quark and a constituent antiquark and has successfully provided a unified description of meson mass spectra
as well as various observables related to their wave functions~\cite{CJ99,CJ99B,HJZR,Dhiman,Arifi22,Arifi:2022qnd,CCH97,Jaus90,Jaus91,C21,Choi21,Arifi:2024mff,Ridwan:2024ngc,Pandya:2024qoj}.
Nevertheless, the LF zero mode remains a long-standing issue in LFD, presenting a major challenge in this framework~\cite{Keister,CJ98,DSHwang,Jaus99,Carbonell98,deMelo:02,BCJ01,BCJ02}.
While leading-twist DAs can typically be computed in a zero-mode-free manner by selecting a ``good" current component, such as $J^+$, 
the calculation of higher-twist DAs is significantly more challenging.
This is because they often involve the ``bad" current component, such as $J^-$, which requires careful treatment of LF zero modes.
Therefore, accurate computations of higher-twist DAs not only serve as a crucial test of the self-consistency of phenomenological models but also provide deeper insights into hadronic quark-gluon dynamics.

In the development of the LFQM, both (1) consistency with manifest covariance and (2) a realistic phenomenological description of hadron structure and spectroscopy have been  key ingredients. 
(1) To ensure the manifest covariance of the LFQM, one may take the approach 
of constructing the LFQM starting from the LF projection of a manifestly covariant Bethe-Salpeter (BS) field-theoretic model.
This approach~\cite{deMelo:02,BCJ01,BCJ02,deMelo:1997cb}, denoted as the ``LFBS model", 
employs Pauli-Villars regularization to extract LF wave functions (LFWFs) and 
typically adopts a symmetric or non-symmetric multipole ansatz for the meson-quark vertex function. 
In this model, the constituent quark ($Q$) and antiquark (${\bar Q}$) are generally off-mass-shell, leading to
non-conservation of LF energy at the meson-quark vertex. This results in a nonzero binding energy, implying $M\neq M_0$, 
where $M$ is the physical rest mass of the meson and $M_0$ is the invariant mass of the quark and antiquark. 
The LF zero mode arising in the LFBS model with a multipole-type wave function
proportional to ${1}/{(M^2 - M^2_0)^n}$, can be
explicitly identified, as discussed in Refs.~\cite{deMelo:1997cb,deMelo:02,BCJ01,BCJ02}. 
Additionally, both potential LF zero modes and instantaneous contributions from the off-shellness of quark propagators can be carefully addressed in this framework.
(2) For the more realistic phenomenology of the LFQM, the approach often referred to
as ``standard LFQM" has been taken,
constructing the model based on a noninteracting quark-antiquark representation.
This approach~\cite{CJ99,CJ99B,HJZR,Dhiman,Arifi:2022qnd,Jaus90,Jaus91,CCH97,Arifi22} employs the spin-orbit (SO) wave function obtained via the interaction-independent Melosh transformation~\cite{Mel1}, 
together with a Gaussian-type or other phenomenological radial wave function.
In contrast to the LFBS model, the standard LFQM constrains the quark and antiquark to be on 
their respective mass shells, ensuring four-momentum conservation at the meson-quark vertex,
$P=p_Q + p_{\bar Q}$, where $P$ and $p_{Q({\bar Q})}$ denote the momenta of the meson and its constituent quark (antiquark), respectively.
Notably, conservation of LF energy ($P^- = p^-_Q + p^-_{\bar Q}$) at the meson-quark vertex 
necessitates treating the meson mass $M$ as the invariant mass $M_0$. 

Our LFQM~\cite{CJ99,CJ99B,HJZR,Dhiman,Arifi:2022qnd,Arifi22} follows the standard LFQM framework, 
incorporating the quark-antiquark interaction potential $V_{Q{\bar Q}}$ through a QCD-motivated effective Hamiltonian, $H_{Q{\bar Q}}=M_0 + V_{Q{\bar Q}}$. The meson mass spectrum is determined by the eigenvalue equation, $H_{Q{\bar Q}}|\Psi\ra =(M_0 + V_{Q{\bar Q}})|\Psi\ra$. 
For consistency with manifest covariance, however, this formulation aligns with the Bakamjian-Thomas (BT) construction~\cite{BT53,KP91}, in which the meson state consists of a noninteracting $Q{\bar Q}$ representation,
with interactions incorporated into the mass operator,  $M:= M_0 + V_{Q{\bar Q}}$.
This approach ensures the preservation of the Poincar\'{e} group structure and commutation relations.

While standard LFQMs generally offer greater predictive power than LFBS models, 
their lack of manifest covariance complicates the treatment of LF zero modes. 
Hadronic matrix elements are typically computed as
\be\label{Eq:1}
\la P'| {\bar q}\Gamma^\mu q|P\ra ={\cal P}^\mu {\cal F},
\ee 
where ${\cal F}$ represents an observable (e.g, decay constants or form factors), 
and ${\cal P}^\mu$ is the associated Lorentz factor.

In most standard LFQM calculations of Eq.~\eqref{Eq:1}, referred to as the ``conventional LFQM",
the BT construction ($M\to M_0$) is applied only to the matrix element, 
while ${\cal P}^\mu$ is treated as an external kinematic variable, independent of internal momenta:
\be\label{ConLFQM}
{\cal F}=\frac{1}{{\cal P}^\mu}\bra{P'}{\bar q}\Gamma^\mu q\ket{P}_{\rm BT}.
\ee
The subscript ``BT"  highlights that the BT construction is applied only to the matrix element, but not to the Lorentz factor ${\cal P}^\mu$.
It is often observed from Eq.~\eqref{ConLFQM} that when ${\cal F}$ is extracted from a ``good" current component, such as the plus current,
where ${\cal P}^+$ is independent of  $M$, it remains free from LF zero modes. 
However, when computed from the minus current, where ${\cal P}^-$ explicitly depends on $M$,
${\cal F}$ generally exhibits LF zero modes.

In our recent studies on the decay constants of pseudoscalar and vector mesons~\cite{Arifi:2022qnd,Arifi:2023uqc} 
and the pion electromagnetic form factor~\cite{CJ2024:pion} within the standard LFQM, we 
explicitly demonstrated that ensuring current-component independence in ${\cal F}$ 
requires the uniform application of the BT construction. Specifically, 
the substitution $M \to M_0$ must be applied consistently to both the matrix element and the Lorentz factor.
This formulation guarantees self-consistency and eliminates LF zero mode contamination by computing 
\be\label{Eq:2}
{\cal F}=\bra{P'}\frac{{\bar q}\Gamma^\mu q}{{\cal P}^\mu}\ket{P}_{\rm BT}, 
\ee
where the subscript ``BT" indicates that ${\cal P}^\mu$ is evaluated within the integral over internal momenta.
In particular, for the minus current, the physical mass $M$ in ${\cal P}^-$ must be replaced with $M_0$.
This substitution effectively absorbs the LF zero mode, ensuring a well-defined and physically meaningful formalism.

The uniform application of the BT construction in the standard LFQM, as demonstrated in Refs.~\cite{Arifi:2023uqc,CJ2024:pion,Ridwan:2024ngc}, 
was first proposed in Ref.~\cite{C13} for vector meson decay constants. 
This approach originates from the covariant BS model~\cite{BCJ01,BCJ02} and transitions to the standard LFQM 
via a matching condition, which we termed the ``Type II" link (see Eq.~(49) in Ref.~\cite{C13}).
A key aspect of this link is replacing $M$ in matrix element integrands with $M_0$, 
effectively imposing the on-mass-shell condition for the constituent quark and antiquark in the LFQM.

Subsequent studies utilizing this link have explored twist-2 and twist-3 DAs and decay constants for pseudoscalar  mesons~\cite{C15,Choi:2017uos},
as well as semileptonic and rare decays between pseudoscalar mesons~\cite{C21,Choi21}, 
consistently yielding current-component-independent results.
Notably, in Ref.~\cite{Arifi:2022qnd}, the equivalence between results obtained via the Type II link and direct LFQM calculations was explicitly demonstrated, further validating its role in ensuring self-consistency across analyses.
To highlight this systematic approach---aligned with the BT construction 
in computing physical observables---we refer to our framework as the ``self-consistent LFQM".

The purpose of this work is to complete the study of vector meson properties by 
extending our previous self-consistent LFQM analyses~\cite{Arifi:2022qnd,C13} on vector meson decay constants and chiral-even DAs to include chiral-odd DAs up to twist 4. 

For the $\rho$ meson, which has both longitudinal $(h=0)$ and transverse $(h=\pm 1)$ polarizations,
there exist two types of decay constants: $f_{\rho}^{\parallel}$ and $f_{\rho}^{\perp}$, respectively.
We demonstrate that these decay constants, associated with polarization vectors $\epsilon_h$, 
can be consistently obtained from the matrix elements $\la 0|{\bar q}(z){\Gamma} q(-z)|\rho(P,h)\ra$,
where $\Gamma=(\gamma^\mu, \sigma^{\mu\nu},\gamma^\mu\gamma_5,\bm{1})$ for both local ($z^\mu=0$) and nonlocal ($z^\mu\neq 0$) matrix elements, in a process-independent manner.
Of particular importance are the nonlocal matrix elements involving axial-vector and scalar currents, where mixing occurs between  
$f_\rho^{\parallel}$ and $f_\rho^{\perp}$. We show that the correct extraction of these decay constants from the mixing is 
ensured by employing the BT construction.

Furthermore, we examine the structure of chiral-even DAs 
($\phi_{2;\mathrm{V}}^\parallel, \phi_{3;\mathrm{V}}^\parallel, \psi_{3;\mathrm{A}}^\perp, \phi_{4;\mathrm{V}}^\parallel$) 
and chiral-odd DAs  
($\phi_{2;\mathrm{T}}^\perp, \phi_{3;\mathrm{T}}^\perp, \psi_{3;\mathrm{S}}^\parallel, \phi_{4;\mathrm{T}}^\perp$) 
beyond leading twist, along with their $\xi$-moments and Gegenbauer moments.
Their behavior in the chiral limit ($m\to 0$) is also computed, showing consistency with results from QCD sum rules~\cite{Ball:1998sk}.

The paper is organized as follows: 
In Sec.~\ref{sec:lfqm}, we outline the LFQM and describe the light-front wave functions of the $\rho(1S)$ meson. 
Section~\ref{sec:const} examines the decay constants derived from four distinct current operators, demonstrating their process independence and rotational invariance. 
In Sec.~\ref{sec:das}, we discuss chiral-even and chiral-odd DAs up to twist 4, obtained from three local and nonlocal current operators $\Gamma$, 
and analyze their $\xi$-moments and Gegenbauer moments. Finally, Section~\ref{sec:summary} presents our conclusions.

\section{Light-front quark model}
\label{sec:lfqm}

In our LFQM~\cite{CJ99,CJ99B,HJZR,Dhiman,Arifi:2022qnd,Arifi22}, a meson is described as a quark-antiquark bound state 
with total momentum $P$, where the Fock state expansion is truncated to include only the constituent quark ($Q$) and antiquark (${\bar Q}$). 
The model represents the Fock state as a noninteracting
$Q{\bar Q}$ pair, with their interactions incorporated into the mass operator, defined as $M:= M_0 + V_{Q{\bar Q}}$.
This formulation ensures consistency with the Poincar\'{e} group's commutation relations for a two-body bound state. 
The resulting interactions are encapsulated in the LFWF $\Psi_{Q{\bar Q}}$, which is obtained as the eigenfunction 
of the QCD-motivated effective Hamiltonian:
$H_{Q{\bar Q}}|\Psi_{Q{\bar Q}}\ra = M |\Psi_{Q{\bar Q}}\ra$~\cite{CJ99,CJ99B,HJZR,Dhiman,Arifi:2022qnd,Arifi22}.

In our LFQM mass spectrum analysis, a Gaussian-type radial wave function~\cite{CJ99,CJ99B} is used as a variational trial function to determine mass eigenvalues and model parameters.
Other types of radial wave functions, such as power-law~\cite{Hwang10}, harmonic oscillator (HO)~\cite{Cardarelli:1994yq,HJZR,Dhiman,Arifi22}, or Gaussian expansion~\cite{Arifi:2024mff}, have also been employed to model the wave functions and improve specific features.
Once these parameters are fixed, they are applied to compute observables such as decay constants and form factors~\cite{CJ99,CJ99B,HJZR,Dhiman,Arifi22,CJ07,C21}.
Focusing on the self-consistent analysis of decay constants and higher-twist DAs for the $\rho(1S)$ meson, we provide a brief overview of the LFWFs, 
highlighting key aspects constrained by the on-mass-shell condition of its constituents.

The meson's four-momentum in LFD is given by
$P=(P^+, P^-, \bm{P}_\perp)$, where $P^+=P^0 + P^3$ and
$P^-=P^0 -P^3$ denote the LF longitudinal momentum and energy, respectively, while $\bm{P}_\perp=(P^1, P^2)$
represents the transverse momenta. We adopt the metric convention $P^2 = P^+P^- - \bm{P}^2_\perp$.

The meson state $|\mathcal{M}(P, J, h)\rangle$, with momentum $P$ and spin state
$(J, h)$, is constructed as~\cite{CCH97,Cheng04,Cheng:2004ew}
\begin{eqnarray}\label{eq:wavefunction}
\ket{\mathcal{M}}
&=& \int \left[ {\rm d}^3{\bar p}_1 \right] \left[ {\rm d}^3{\bar p}_2 \right]  2(2\pi)^3 \delta^3 \left({\bar P}-{\bar p}_1-{\bar p}_2 \right)\nonumber\\
&&\times 
\sum_{\lambda_1,\lambda_2} \Psi_{\lambda_1 \lambda_2}^{Jh}(x, \bm{k}_\perp)
\ket{q_{\lambda_1}(p_1) \bar{q}_{\lambda_2}(p_2) },
\end{eqnarray}
where $p^\mu_i$ and $\lambda_i$ are the on-mass-shell ($p^2_i=m^2_i$) momenta and helicities of the
constituent quark ($i=1$) and antiquark $(i=2)$, respectively.
The LF three-momentum is defined as ${\bar p}=(p^+,\bm{p}_\perp)$, with the integration measure
$\left[ {\rm d}^3{\bar p} \right] \equiv {\rm d}p^+ {\rm d}^2\bm{p}_{\perp}/(16\pi^3)$.
The LF internal relative momentum variables $(x, \bm{k}_\perp)$ are
defined as $x_i=p^+_i/P^+$ and $\bm{k}_{i\perp} = \bm{p}_{i\perp} - x_i \bm{P}_\perp$, where  
$\sum_i x_i=1$ and $\sum_i\bm{k}_{i\perp}=0$. We set $x=x_1$ and $\bm{k}_\perp =\bm{k}_{1\perp}$.
The meson state satisfies the normalization condition:
\bea\label{eq:norm1}
\la{\mathcal{M}^\prime}|\mathcal{M}\ra = 2 (2\pi)^3 P^+ \delta^3({\bar P}^\prime -{\bar P})\delta_{J' J}\delta_{h^\prime h},
\eea
where $\mathcal{M}^\prime = \mathcal{M}(P', J', h')$.

In momentum space, the LFWF  of a meson is decomposed as
\begin{eqnarray}\label{eq:LFWF}
	\Psi^{Jh}_{\lambda_1\lambda_2} (x, \bm{k}_{\bot}) &=& \Phi(x, \bm{k}_\bot) \mathcal{R}^{Jh}_{\lambda_1\lambda_2}(x, \bm{k}_\bot),
\end{eqnarray}
where $\Phi(x, \bm{k}_\bot)$ is the radial wave function used in our variational mass spectroscopic analysis~\cite{CJ99,CJ99B,HJZR,Dhiman,Arifi22}, and $\mathcal{R}^{Jh}_{\lambda_1\lambda_2 }$ 
is the spin-orbit (SO) wave function, obtained via the interaction-independent Melosh transformation~\cite{Mel1},
from the ordinary SO wave function assigned by the quantum number $J^{PC}$.

A key feature of the LF formulation for bound states, as expressed in Eq.~\eqref{eq:wavefunction}, is the frame independence of the LFWF~\cite{BPP}. 
Unlike in the instant form, where wave functions depend on the hadron's overall motion, the internal variables $(x, \bm{k}_\perp)$ in the LFWF
remain unchanged under boosts to any physical $(P^+, \bm{P}_\perp)$ frame. 

In our LFQM, the $\rho$ meson is treated under SU(2) isospin symmetry, assuming equal up and down quark masses, 
denoted as $m_1=m_2=m$.
The interaction-independent SO wave function, consistent with the BT construction,
can then be written in a covariant form~\cite{Jaus90,Jaus91} as
\begin{equation} \label{eq:spinorbit}
	\mathcal{R}^{1h}_{\lambda_1\lambda_2} 
    = \frac{1}{\sqrt{2} M_0}\bar{u}_{\lambda_1}^{}(p_1) \left[-\slashed{\epsilon}_h + \frac{(p_1-p_2)\cdot{\epsilon}_h}{M_0 + 2m}\right] v_{\lambda_2}^{}(p_2),
\end{equation}
where the boost-invariant meson mass squared is given by
\begin{eqnarray}
M_0^2 = \frac{\bm{k}_{\bot}^2 + m^2}{x(1-x)}.
\end{eqnarray}
In this LFQM framework, based on the BT construction, 
the polarization vectors ${\epsilon}^\mu_h$ $(h=\pm 1, 0)$  satisfy the transversality condition
$\epsilon\cdot P=0$, where $P^-=\frac{M^2_0 + {\bf P}^2_\perp}{P^+}$ ensures the LF energy conservation,
i.e., $P^-=p^-_Q + p^-_{\bar Q}$, at the meson-quark vertex. 
Consequently, the polarization vectors are expressed in terms of $M_0$ as~\cite{Jaus90,Jaus91}
\begin{equation}
\begin{aligned}
{\epsilon}^\mu_\pm &= \left( 0, \frac{2}{P^+} \boldsymbol{\epsilon}_\perp(\pm) \cdot \bm{P}_\perp, \boldsymbol{\epsilon}_\perp(\pm)\right), \\
{\epsilon}^\mu_0&= \frac{1}{M_0}\left(P^+, \frac{ \bm{P}^2_\perp-M^2_0 }{P^+}, \bm{P}_\perp\right),
\end{aligned}
\end{equation}
where $\boldsymbol{\epsilon}_\perp(\pm)= \mp \frac{1}{\sqrt{2}}\left( 1, \pm i \right)$. 

Using Dirac helicity spinors~\cite{Lepage:1980fj,Jaus90} along with the polarization vectors above,
the explicit matrix forms of the SO wave functions for the vector meson are given by
\begin{eqnarray}\label{VSO}
\mathcal{R}^{1+1}_{\lambda_1 \lambda_2} (x,\bm{k}_\perp)  &=& \mathcal{R}_0
\begin{pmatrix}
m + \frac{{\bm k}_\bot^2}{M_0 + 2m} & k_\perp^{(+)} \frac{\mathcal{M}_1}{M_0 + 2m}\\
- k_\perp^{(+)} \frac{\mathcal{M}_2}{M_0 + 2m}    & -\frac{(k_\perp^{(+)})^2}{M_0 + 2m} 	\\
\end{pmatrix},
\nonumber\\
\mathcal{R}^{10}_{\lambda_1 \lambda_2} (x,\bm{k}_\perp)  &=& \frac{\mathcal{R}_0}{\sqrt{2}}
\begin{pmatrix}
k_\perp^{(-)} \frac{\mathcal{M}_2-\mathcal{M}_1}{M_0 + 2m} & m + \frac{2{\bm k}_\bot^2}{M_0 + 2m} \\
m + \frac{2{\bm k}_\bot^2}{M_0 + 2m}   &   -k_\perp^{(+)} \frac{\mathcal{M}_2-\mathcal{M}_1}{M_0 + 2m} \\
\end{pmatrix},\\
\mathcal{R}^{1-1}_{\lambda_1 \lambda_2} (x,\bm{k}_\perp)  &=& \mathcal{R}_0
\begin{pmatrix}
-\frac{(k_\perp^{(-)})^2}{M_0 + 2m} & k_\perp^{(-)} \frac{\mathcal{M}_2}{M_0 + 2m}\\
- k_\perp^{(-)} \frac{\mathcal{M}_1}{M_0 + 2m}    &   m + \frac{{\bm k}_\bot^2}{M_0 + 2m} 	\\
\end{pmatrix}, \nonumber\qquad
\end{eqnarray}
where $k_\perp^{(\pm)}=k_x\pm i k_y$,
$\mathcal{R}_0 =1/\sqrt{m^2 + \bm{k}_\perp^2}$, and
$ \mathcal{M}=\mathcal{M}_2-\mathcal{M}_1$ with
$\mathcal{M}_1= x M_0 + m$ and $\mathcal{M}_2 = (1 - x) M_0 + m.$ 
We note that  $\mathcal{R}^{Jh}_{\lambda_1\lambda_2}$ satisfy the orthogonality condition
$\sum_{\lambda_1\lambda_2} \mathcal{R}^{Jh\dagger}_{\lambda_1\lambda_2} \mathcal{R}^{J^\prime h\prime}_{\lambda_1\lambda_2} = \delta_{JJ^\prime}\delta_{hh^\prime}$.

For the radial wave function, we adopt the following rotationally invariant Gaussian form:
\begin{equation}\label{GWk2}
	\Phi (\bm{k}) = \frac{4\pi^{3/4}}{\beta^{3/2}} e^{-\bm{k}^2/ 2\beta^2},
\end{equation}
where $\bm{k}=(k_z, \bm{k}_\perp)$, and $\beta$ serves as a variational parameter in our mass spectroscopic 
analysis~\cite{CJ99,CJ99B,HJZR,Dhiman,Arifi22}.
The Gaussian wave function satisfies the normalization condition
\begin{equation}
\int\; \frac{{\rm d}^3\bm{k}}{2(2\pi)^3}\;\abs{ \Phi(\bm{k}) }^2 =1.
\end{equation}
To ensure rotational invariance, 
we express the normalization of $\Phi(\bm{k})$ in terms of $\Phi (x, \bm{k}_\perp)$ by transforming variables $(k_z, \bm{k}_\perp) \to (x,\bm{k}_\perp)$:
\begin{equation}\label{eq:norm2}
 \int_0^1  {\rm d}x \int \frac{{\rm d}^2 \bm{k}_\bot}{2(2\pi)^3}  \abs{ \Phi(x, \bm{k}_\bot) }^2 =1.
\end{equation} 
The wave function $\Phi(x, \bm{k}_\bot)$ includes the Jacobian factor:
\begin{equation}\label{HO1S2SJac}
    \Phi(x,\bm{k}_\perp) = \sqrt{\frac{\partial k_z}{\partial x}}\Phi(\bm{k}),
\end{equation}
where $k_z = \frac{1}{2}(2x-1) M_0$ and $\frac{\partial k_z}{\partial x} = \frac{M_0}{4x(1-x)}$
for the case of equal quark and antiquark masses.

The total LFWF, given by Eq.~\eqref{eq:LFWF}, satisfies the same normalization condition as Eq.~\eqref{eq:norm2}. 
This follows from the meson state $|\mathcal{M}(P, J, h)\rangle$ obeying the normalization in Eq.~\eqref{eq:norm1}, 
and the SO wave function being the unitary. 
In particular, the inclusion of the Jacobian factor in $\Phi(x,\bm{k}_\perp)$ ensures the rotational invariance of the model wave function,
yielding self-consistent results that are independent of current components and invariant under boosts.

\section{Decay constants}
\label{sec:const}

In this section, we begin by reviewing the calculation of decay constants derived from local matrix elements of the vector and tensor current operators, using various current components and polarization vectors~\cite{Arifi:2022qnd}, as summarized in Table~\ref{tab:t1}.
We then provide a detailed analysis of decay constants obtained from nonlocal matrix elements involving axial-vector and scalar current operators, which is the main focus of this work.
Particular emphasis is placed on achieving self-consistency within the model, based on the BT construction~\cite{C13,C15,Choi:2017uos,Arifi:2022qnd,C21}.

\subsection{Vector and Tensor Currents}

The decay constants $f_{\rho}^{\parallel}$ and $f_{\rho}^{\perp}$, corresponding to the 
 longitudinally and transversely polarized $\rho$ meson,
are defined via the local matrix elements of the vector ($\Gamma^\mu_\mathrm{V}\equiv\gamma^\mu$) and tensor ($\Gamma^{\mu\nu}_{\rm T}\equiv\sigma^{\mu\nu}$) current operators~\cite{Ball:1998sk}

\begin{eqnarray} \label{eq:local1}
	\bra{0} \bar{q}(0) \Gamma^\mu_\mathrm{V} q(0) \ket{{\rho}(P,h)} &=& f_{\rho}^\parallel \mathcal{P}^\mu_{\rm V},
    \nonumber\\
 	\bra{0} \bar{q}(0) \Gamma^{\mu\nu}_{\rm T} q(0) \ket{{\rho}(P,h)} &=& f_{\rho}^\perp \mathcal{P}^{\mu\nu}_{\rm T},
	 \label{eq:local2}
\end{eqnarray}
where $\mathcal{P}^\mu_{\rm V}= M \epsilon^\mu_h$ and $\mathcal{P}^{\mu\nu}_{\rm T}=i \left(\epsilon^\mu_h P^\nu -\epsilon^\nu_h P^\mu \right)$.
Here, $M$ and $\epsilon^\mu_h$ denote the mass and polarization vector of the $\rho$ meson. 

In the standard LFQM, the decay amplitudes for the current operators 
$\Gamma=\Gamma^\mu_\mathrm{V}$ or $\Gamma= \Gamma^{\mu\nu}_{\rm T}$ are given by
\begin{eqnarray}\label{eq:slf0}
\bra{0} \bar{q}\Gamma q \ket{\rho(P,h)} &=& \sqrt{N_c} \int_0^1 {\rm d}x \int \frac{ {\rm d}^2 \bm{k}_\bot}{16\pi^3}\  \Phi(x,\bm{k}_\perp)  \nonumber\\
   & &  \times \mbox{}
   \sum_{\lambda_1, \lambda_2} \mathcal{R}_{\lambda_1 \lambda_2}^{1h} \left[\frac{\bar{v}_{\lambda_2}(p_2)}{\sqrt{x_2}} \Gamma  \frac{u_{\lambda_1}(p_1)}{\sqrt{x_1}}\right],\quad \quad 
\end{eqnarray}
where the color factor $N_c=3$ arises from the implicit color structure of the wave function~\cite{Jaus91,CCH97}.

In the LFQM analysis~\cite{Arifi:2022qnd}, formulated within the BT construction $(M\to M_0)$, 
we explicitly demonstrated that $f_{\rho}^{\parallel}$ and $f_{\rho}^{\perp}$ can be obtained in a way 
that is independent of current components, polarization vectors, and reference frames,
by computing
\begin{eqnarray}\label{fVfTQM}
f_{\rho}^\parallel &=& \bra{0}\frac{{\bar q}\gamma^\mu q}{\mathcal{P}^\mu_{\rm V}}\ket{P}_{\rm BT}, \nonumber\\
f_{\rho}^\perp &=& \bra{0}\frac{{\bar q}\sigma^{\mu\nu} q}{\mathcal{P}^{\mu\nu}_{\rm T}}\ket{P}_{\rm BT}.
\end{eqnarray}
Here, the physical mass $M$ included in Lorentz factors $\mathcal{P}^\mu_{\rm V}$ and $\mathcal{P}^{\mu\nu}_{\rm T}$
is replaced with the invariant mass $M_0$,
and the Lorentz factors are evaluated within the integral over internal momenta.

\begin{table}[t]
	\begin{ruledtabular}
		\renewcommand{\arraystretch}{1.5}
		\caption{Operators $\mathcal{O}_{\rm V(T)}$ from Eq.~\eqref{eq:OPVT} for all possible components of the currents
        $\Gamma^\mu_{\rm V}(\Gamma^{\mu\nu}_{\rm T})$ and the polarization vectors $\epsilon_h$~\cite{Arifi:2022qnd}.}
		\label{tab:t1}
		\begin{tabular}{cccc}
			 & $\Gamma_{\rm V}=\gamma^\mu$ & \hspace{0.2cm} $\epsilon_h$ \hspace{0.2cm} & $\mathcal{O}^\mu_{\rm V}(h)$\\  \hline 
		    \multirow{6}{*}{$f_{\rho}^{\parallel}$}  & \multirow{2}{*}{$\gamma^{+},\gamma^\perp$} 
		    & \multirow{2}{*}{$\epsilon_0$} 
		    & \multirow{2}{*}{$\sqrt{2}{\cal R}_0\left(2m + \frac{4\bm{k}_\bot^2}{M_0 + 2m}\right)$} \\
		    &&&\\
			   & \multirow{2}{*}{$\gamma^-$}   
			   & \multirow{2}{*}{$\epsilon_0$} 
                 & \multirow{2}{*}{$\sqrt{2}{\cal R}_0\left(2m + \frac{4\bm{k}_\bot^2}{M_0 + 2m}\right)$} \\
                &&& \\ 
			    & \multirow{2}{*}{$\gamma^{\perp},\gamma^-$} 
			   & \multirow{2}{*}{$\epsilon_{+}$} 
                & \multirow{2}{*}{$\sqrt{2}{\cal R}_0\left(M_0 - \frac{2\bm{k}_\bot^2}{M_0 + 2m}\right)$}  \\ 
                &&&  \\ \cline{1-4}
                & $\Gamma_{\rm T}=\sigma^{\mu\nu}$ & \hspace{0.2cm} $\epsilon_h$ \hspace{0.2cm} & $\mathcal{O}^{\mu\nu}_{\rm T}(h)$\\  \hline
            \multirow{6}{*}{$f_{\rho}^{\perp}$} & \multirow{2}{*}{$\sigma^{\perp +},\sigma^{+-}$} & \multirow{2}{*}{$\epsilon_+$}  
            &  \multirow{2}{*}{$\sqrt{2}{\cal R}_0\left(2m + \frac{2\bm{k}_\bot^2}{M_0 + 2m}\right)$} \\  
            &&& \\
            & \multirow{2}{*}{$\sigma^{\perp -}$} & \multirow{2}{*}{$\epsilon_+$} 
            & \multirow{2}{*}{$\sqrt{2}{\cal R}_0 \left(2m + \frac{2\bm{k}_\bot^2}{M_0 + 2m}\right)$}\\ 
            &&& \\
            & \multirow{2}{*}{$\sigma^{+ -},\sigma^{\perp-}$} & \multirow{2}{*}{$\epsilon_0$} 
            & \multirow{2}{*}{$\sqrt{2}{\cal R}_0 \left(M_0 - \frac{4\bm{k}_\perp^2}{M_0 + 2m}\right)$}\\ 
            &&& \\ 
		\end{tabular}
		\renewcommand{\arraystretch}{1}
	\end{ruledtabular}
\end{table}

The decay constants $f_{\rho}^{\parallel(\perp)}$, defined in Eq.~\eqref{fVfTQM}, can be expressed in a generic integral form as
\begin{eqnarray}\label{eq:slf}
f_{\rho}^{\parallel(\perp)} &=& \sqrt{N_c} \int_0^1 {\rm d}x \int \frac{ {\rm d}^2 \bm{k}_\bot}{16\pi^3}\  \Phi(x,\bm{k}_\perp)  \nonumber\\
   & &  \times \mbox{} \frac{1}{\mathcal{P}_{\rm V(T)}}
   \sum_{\lambda_1, \lambda_2} \mathcal{R}_{\lambda_1 \lambda_2}^{1h} \left[\frac{\bar{v}_{\lambda_2}(p_2)}{\sqrt{x_2}}
   \Gamma_{\rm V(T)}  \frac{u_{\lambda_1}(p_1)}{\sqrt{x_1}}\right].
   \nonumber\\
\end{eqnarray}
For simplicity, we omit the explicit Lorentz indices of $(\mathcal{P}^\mu_{\rm V}, \Gamma^\mu_{\rm V})$ 
and $(\mathcal{P}^{\mu\nu}_{\rm T}, \Gamma^{\mu\nu}_{\rm V})$, 
although they are implicitly carried through the calculations
for various current components and polarization vectors.

While the explicit forms of $f_{\rho}^{\parallel(\perp)}$ for all current components and polarization vectors were provided in Ref.~\cite{Arifi:2022qnd},
we summarize the final compact expressions here for completeness:
\be\label{eq:operator}
f_{\rho}^{\parallel(\perp)} = \sqrt{N_c}  \int_0^1 {\rm d}x \int \frac{ {\rm d}^2 \bm{k}_\bot}{16\pi^3}\  
	\Phi(x, \bm{k}_\bot) ~\mathcal{O}_{\rm V(T)},
\ee
where the operators $\mathcal{O}_{\rm V(T)}$ are defined as
\be\label{eq:OPVT}
\mathcal{O}_{\rm V(T)}=\frac{1}{\mathcal{P}_{\rm V(T)}}
   \sum_{\lambda_1, \lambda_2} \mathcal{R}_{\lambda_1 \lambda_2}^{1h} \left[\frac{\bar{v}_{\lambda_2}(p_2)}{\sqrt{x_2}}
   \Gamma_{\rm V(T)}  \frac{u_{\lambda_1}(p_1)}{\sqrt{x_1}}\right].
\ee
These operators are derived using the Dirac helicity spinors~\cite{Lepage:1980fj,Jaus90} and the SO wave functions defined in 
Eq.~\eqref{VSO}. 

The complete results for $f_{\rho}^{\parallel(\perp)}$ with all current components and polarization vectors
are summarized in Table~\ref{tab:t1}. Notably, 
the expressions for $f_{\rho}^{\parallel}$ obtained from $\ep_0$ with any current $\gamma^\mu$ are identical in functional form.
Although the results derived using $\ep_+$ with $(\gamma^\perp,\gamma^-)$ differ in form from those with $\ep_0$,
they yield the same numerical values. This demonstrates the current-component independence of our formalism,
with a similar pattern observed for $f_{\rho}^{\perp}$.
This independence is achieved through the incorporation of the Lorentz structures $\mathcal{P}_{\rm V(T)}$
inside the integral and the consistent replacement of $M$ with $M_0$. These features
ensure that the decay constants are independent of
current components and polarization vectors, maintain boost invariance, and remain self-consistent within the model 
framework---a key hallmark of our LFQM based on the BT construction~\cite{C13,C15,Choi:2017uos,Arifi:2022qnd,C21}. 

\subsection{Axial-Vector and Scalar Currents}

The main focus of this work is the calculation of nonlocal matrix elements of the
axial-vector ($\Gamma^{\mu}_{\rm A}=\gamma^\mu\gamma_5$) and scalar ($\Gamma_{\rm S}=\bm{1}$) current operators, further extending our LFQM
framework based on the BT construction~\cite{C13,C15,Choi:2017uos,Arifi:2022qnd,C21}.

The decay constants of the $\rho$ meson can be extracted from these nonlocal matrix elements, defined in Ref.~\cite{Ball:1998sk}, as
\begin{eqnarray} \label{eq:nonlocal1}
A_\mathrm{A}^\mu &=& \langle 0| \bar{q}(z)\gamma^\mu\gamma^5 [z,-z]q(-z)|\rho(P,h) \rangle \nonumber \\
&=& f_\rho^\mathrm{A}{\cal P}^\mu_{\rm A}\int^1_0 \dd x\ e^{i\zeta P\cdot z}  \psi_{3;{\rm A}}^\perp(x),   
\quad\quad \nonumber\\
A_\mathrm{S} &=& \langle 0| \bar{q}(z) \bm{1} [z,-z] q(-z)|\rho(P,h) \rangle \nonumber \\
&=& -if_\rho^\mathrm{S}{\cal P}_{\rm S} \int^1_0 \dd x\ e^{i\zeta P\cdot z} \psi_{3;{\rm S}}^\parallel(x),\label{eq:nonlocal2}
\end{eqnarray}
where ${\cal P}^\mu_{\rm A}=\frac{1}{2}M \epsilon^\mu_{\nu\alpha\beta} \epsilon^\nu_h P^\alpha z^\beta$,
${\cal P}_{\rm S}= M^2 (\epsilon_h \cdot z)$, and $\zeta=2x-1$. 
The constants $f_\rho^\mathrm{A}$ and $f_\rho^\mathrm{S}$ are the decay constants associated with the axial-vector and scalar currents, 
respectively. The two-particle twist-3 DAs, $\psi_{3;\mathrm{A}}^\perp(x)$ and $\psi_{3;\mathrm{S}}^\parallel(x)$,
satisfy the normalization condition
\begin{equation}\label{eq:DAnorm}
\int^1_0\;\dd x\;\psi_{3;\mathrm{A(S)}}^{\perp(\parallel)}(x)=1.
\end{equation}
The path-ordered gauge factor $[z,-z]$, ensuring gauge invariance, is set to unity in the LF gauge ($A^+=0$) throughout this work.
Since $f^{\rm A}_\rho$ and $f^{\rm S}_\rho$ are defined 
through these nonlocal matrix elements, their extraction requires working
in the nonlocal LF limit, with $z^+=\bm{z}_\perp=0$. 
Thus, determining $\psi_{3;{\rm A}}^\perp(x)$ and $\psi_{3;{\rm S}}^\parallel(x)$ 
is essential to obtain these decay constants.

We first present the calculation of $\psi_{3;{\rm S}}^\parallel(x)$ and $f^{\rm S}_\rho$
by integrating both sides of Eq.~(\ref{eq:nonlocal2}) with respect to $z^-$,
 using the dummy variable $x^\prime$ (and $\zeta'=2x'-1$), as follows:
\begin{eqnarray} \label{eq:ten1}
&&\int_{-\infty}^\infty \frac{{\rm d}z^-}{2\pi} {\rm e}^{-i\zeta^\prime P\cdot z}  
\bra{0} \bar{q}(z) q(-z)\ket{\rho(P,h)} \nonumber\\
&&= C_S \int_0^1 {\rm d}x\ \int_{-\infty}^\infty \frac{{\rm d}z^-}{2\pi} z^- {\rm e}^{-i(x^\prime-x) P^+z^-}  \psi_{3;{\rm S}}^\parallel(x),\quad \quad 
\end{eqnarray}
where $C_S=-\frac{i}{2}f_\rho^{\rm S}M^2\epsilon^+_h$. 
It is important to note that only the longitudinal ($h=0$) polarization vector contributes 
to the extraction of $f_\rho^{\rm S}$.

The integration of the right-hand side (RHS) of Eq.~\eqref{eq:ten1} over $z^-$ 
is given by~\cite{Arifi:2023uqc}
\bea
&&\int_{-\infty}^\infty \frac{{\rm d}z^-}{2\pi} z^- {\rm e}^{-i(x^\prime-x) P^+z^-}  \psi_{3;{\rm S}}^\parallel(x)
\nonumber\\
&&= \frac{i}{(P^+)^2}\frac{\partial }{\partial x'}
\left[\delta(x-x')\psi_{3;{\rm S}}^\parallel(x)\right]. \quad \quad 
\eea
This leads to
\be\label{eq:RHS}
{\rm RHS\; of\; Eq.~\eqref{eq:ten1} } =
\frac{f_\rho^\mathrm{S} M^2\epsilon^+_0}{2(P^+)^2} \frac{\partial \psi_{3;{\rm S}}^\parallel(x')}{\partial x'}.
\ee
The left-hand side (LHS) of Eq.~\eqref{eq:ten1} can be written in our LFQM as
\bea\label{eq:LHS}
&&{\rm LHS\; of\; Eq.~\eqref{eq:ten1} } \nonumber\\
&&=  \sqrt{N_c}\int^1_0 {\rm d}x\int \frac{ {\rm d}^2\bm{k}_\perp}{16\pi^3} 
\int_{-\infty}^\infty \frac{{\rm d}z^-}{2\pi} {\rm e}^{-i\zeta^\prime P\cdot z} {\rm e}^{-i(p_2-p_1)\cdot z}
\nonumber\\
&&\times 
\sum_{\lambda_1, \lambda_2}
   \Psi_{\lambda_1 \lambda_2}^{10}(x, \bm{k}_\perp)
  \left[\frac{\bar{v}_{\lambda_2}(p_2)}{\sqrt{x_2}} \frac{u_{\lambda_1}(p_1)}{\sqrt{x_1}}\right],
\end{eqnarray}
where $ \Psi_{\lambda_1 \lambda_2}^{10}(x, \bm{k}_\perp)$ is the LFWF defined in Eq.~\eqref{eq:LFWF}.
Since ${\rm e}^{-i\zeta^\prime P\cdot z} {\rm e}^{-i(p_2-p_1)\cdot z}={\rm e}^{-i(x'-x)P^+z^-}$, 
the $z^-$ integration yields $\delta[(x^\prime - x)P^+]$. Substituting this result leads to
\bea\label{eq:LHS1}
&&{\rm LHS\; of\; Eq.~\eqref{eq:ten1} } \nonumber\\
&&=  \frac{\sqrt{N_c}}{P^+}\int \frac{ {\rm d}^2\bm{k}_\perp}{16\pi^3}  
\sum_{\lambda_1, \lambda_2}
   \Psi_{\lambda_1 \lambda_2}^{10}(x', \bm{k}_\perp)
  \left[\frac{\bar{v}_{\lambda_2}(p'_2)}{\sqrt{x'_2}} \frac{u_{\lambda_1}(p'_1)}{\sqrt{x'_1}}\right],
  \nonumber\\
\end{eqnarray}
where the prime (\('\)) in  \(\Psi_{\lambda_1 \lambda_2}^{10}(x', \bm{k}_\perp)\) and in \(p_i\) 
indicates that the longitudinal momentum fractions inside the integral are expressed in terms of \(x'\), 
specifically, \(x'_1 = x'\) and \(x'_2 = 1 - x'\).

By integrating both Eqs.~\eqref{eq:RHS} and~\eqref{eq:LHS1} over $x'$, we obtain 
\begin{eqnarray}\label{eq:psi3_s}
\psi_{3;{\rm S}}^\parallel(x) &=& \frac{\sqrt{N_c}}{f^{\rm S}_{\rho}}
\int_0^x {\rm d}x^\prime \int \frac{ {\rm d}^2\bm{k}_\perp}{16\pi^3} \Phi(x^\prime,\bm{k}_\perp) \nonumber\\ 
&& \times \frac{2P^+}{M^2\epsilon^+_0} \sum_{\lambda_1, \lambda_2}
   \mathcal{R'}_{\lambda_1 \lambda_2}^{10} 
  \left[\frac{\bar{v}_{\lambda_2}(p'_2)}{\sqrt{x_2^\prime}}  \frac{u_{\lambda_1}(p'_1)}{\sqrt{x_1^\prime}}\right]. \quad \quad
\end{eqnarray}
By integrating both sides over $x$ and applying the normalization condition given in Eq.~\eqref{eq:DAnorm}, 
the decay constant $f^{{\rm S}}_\rho$ is given by
\begin{eqnarray}\label{eq:f_scalar}
f_{\rho}^\mathrm{S} &=&  \sqrt{N_c} \int_0^1 {\rm d}x \int_0^x {\rm d}x^\prime \int \frac{ \dd^2\bm{k}_\perp}{16\pi^3}  \Phi(x^\prime,\bm{k}_\perp) \nonumber\\
&& \times \frac{2P^+}{M^2\epsilon^+_0} \sum_{\lambda_1, \lambda_2}
   \mathcal{R'}_{\lambda_1 \lambda_2}^{10}
  \left[\frac{\bar{v}_{\lambda_2}(p'_2)}{\sqrt{x_2^\prime}}  \frac{u_{\lambda_1}(p'_1)}{\sqrt{x_1^\prime}}\right].\quad 
\end{eqnarray} 
In our LFQM based on the BT construction, we replace the physical mass $M$ with the invariant mass $M_0$ 
in the integrand,
i.e., $\frac{P^+}{M^2\epsilon^+_0}\to\frac{1}{M_0}$, to obtain $f_{\rho}^\mathrm{S}$. 
This yields the final expression:
\be\label{eq:f_scalar2}
f_{\rho}^\mathrm{S} =  \sqrt{N_c} \int_0^1 {\rm d}x \int_0^x {\rm d}x^\prime \int \frac{ \dd^2\bm{k}_\perp}{16\pi^3}  \Phi(x^\prime,\bm{k}_\perp)\; {\cal O}'_{\rm S},
\ee
where
\be\label{eq:Os}
{\cal O}'_{\rm S}=2\sqrt{2}{\cal R}_0 (1-2x') M'_0.
\ee
The resulting numerical value of $f_{\rho}^\mathrm{S}$ is then substituted back into Eq.~\eqref{eq:psi3_s}, applying the same replacement 
to determine $\psi_{3;{\rm S}}^\parallel(x)$. 

Now, for the computation of $\psi_{3;{\rm A}}^\perp(x)$ and $f^{\rm A}_\rho$ in Eq.~\eqref{eq:nonlocal1},
the only nonvanishing Lorentz factor is 
${\cal P}^\mu_{\rm A}=\frac{1}{2}M \epsilon^\mu_{\nu\alpha -} \epsilon^\nu_\pm p^\alpha z^-$. 
This implies that only the transverse ($h=\pm$) polarization vectors contribute. 
Furthermore, in the $\bm{P}_\perp=0$ frame, only ${\cal P}^\perp_{\rm A}$ remains nonzero. 
For instance\footnote{In this work, we adopt the convention $\epsilon^{+-xy}=2$ and use LF metric components
$g^{+-}=g^{-+}=2$}., 
for $\mu=x$, $\epsilon^x_{y+-} \epsilon^y_+ p^+ z^-= \frac{-iP^+z^-}{2\sqrt{2}}$; and for $\mu=y$, $\epsilon^y_{x+-} \epsilon^x_+ p^+ z^-= \frac{P^+z^-}{2\sqrt{2}}$. 
The resulting expressions for $\psi_{3;\mathrm{A}}^\perp(x)$ and $f_{\rho}^\mathrm{A}$ 
are independent of the choice of polarization vector $(h=\pm)$ and current component $(\mu=x,y)$.

Following the same procedure used for the scalar current and choosing the combination $(h=+, \mu=x)$, we obtain the analogs of
Eqs.~\eqref{eq:psi3_s} and~\eqref{eq:f_scalar} for $\psi_{3;\mathrm{A}}^\perp(x)$ and $f_{\rho}^\mathrm{A}$:
\begin{eqnarray}\label{eq:psi3_a}
\psi_{3;\mathrm{A}}^\perp(x) &=& 
\frac{\sqrt{N_c}}{f^{\rm A}_{\rho}} \int_0^x {\rm d}x^\prime \int \frac{ {\rm d}^2\bm{k}_\perp}{16\pi^3} \Phi(x^\prime,\bm{k}_\perp) \nonumber\\ 
&& \times \frac{4\sqrt{2}}{M} \sum_{\lambda_1, \lambda_2}
   \mathcal{R'}_{\lambda_1 \lambda_2}^{11} 
  \left[\frac{\bar{v}_{\lambda_2}(p'_2)}{\sqrt{x_2^\prime}}\gamma^x\gamma_5 \frac{u_{\lambda_1}(p'_1)}{\sqrt{x_1^\prime}}\right], \quad \quad
\end{eqnarray}
and
\begin{eqnarray}\label{eq:f_axialvector}
f_{\rho}^\mathrm{A} &=&  \sqrt{N_c} \int_0^1 {\rm d}x \int_0^x {\rm d}x^\prime \int \frac{ \dd^2\bm{k}_\perp}{16\pi^3}  \Phi(x^\prime,\bm{k}_\perp) \nonumber\\
&& \times \frac{4\sqrt{2}}{M} \sum_{\lambda_1, \lambda_2}
   \mathcal{R'}_{\lambda_1 \lambda_2}^{11}
  \left[\frac{\bar{v}_{\lambda_2}(p'_2)}{\sqrt{x_2^\prime}} \gamma^x\gamma_5 \frac{u_{\lambda_1}(p'_1)}{\sqrt{x_1^\prime}}\right].\quad 
\end{eqnarray} 
Finally, following the same approach used for $f_{\rho}^\mathrm{S}$, we obtain 
\be\label{eq:f_scalar2}
f_{\rho}^\mathrm{A} =  \sqrt{N_c} \int_0^1 {\rm d}x \int_0^x {\rm d}x^\prime \int \frac{ \dd^2\bm{k}_\perp}{16\pi^3}  \Phi(x^\prime,\bm{k}_\perp) \;{\cal O}'_{\rm A},
\ee
where ${\cal O}'_{\rm A}=2{\cal O}'_{\rm S}$, demonstrating that $f^{\rm A}_\rho = 2 f^{\rm S}_\rho$ in our model calculation.

Using the QCD equation of motion, it was shown in Ref.~\cite{Ball:1998sk} that the decay constants
$f^{\rm A}_\rho$ and $f^{\rm S}_\rho$ can be expressed as linear combinations of $f_{\rho}^\parallel$ and $f_{\rho}^\perp$:
\begin{eqnarray}\label{eq:fAfS}
    f_\rho^\mathrm{A} &=& f_\rho^\parallel- \alpha f_\rho^\perp,\nonumber\\
    f_\rho^\mathrm{S} &=& f_\rho^\perp - \alpha f_\rho^\parallel.
\end{eqnarray}
where $\alpha=2m/M$ accounts for quark mass corrections.
Here, the quark mass $m$ is understood as a constituent quark mass that effectively encodes non-perturbative dynamics 
such as chiral symmetry breaking and confinement at the hadronic scale. It does not correspond to the QCD scale-dependent running mass. 
In QCD, the relation $\alpha = 2m/M$, where $M$ is derived from the equation-of-motion relations among light-cone DAs.
However, in our BT-based LFQM, the meson is described as a composite \( q\bar{q} \) system characterized by an internal invariant mass $M_0$, 
which differs from the physical mass $M$. To ensure that Eq.~(35) is satisfied self-consistently within our model, 
we should take $M$ as $M_0$, yielding $\alpha = 2m/M_0$. This assignment is not arbitrary but essential for obtaining decay constants and DAs that are independent of the choice of current component or polarization state. In contrast, employing a fixed value of  $\alpha = 2m/M$ in our model would lead
to inconsistencies across different current components. Thus, our consistent treatment of $M$ in the entire computation ensures internal 
consistency and  effectively implements
the QCD equation-of-motion constraint within the LF framework.

To verify this consistency explicitly, we demonstrate that Eq.~\eqref{eq:fAfS} also holds within our LFQM framework.
To maintain consistency with the BT construction, the physical mass $M$ in $\alpha$ must be replaced by the invariant mass $M_0$,
and this replacement must be applied within the integrals. 
Thus, Eq.~\eqref{eq:fAfS} becomes, in our LFQM:
\begin{eqnarray}\label{eq:fAfSQM}
    f_\rho^\mathrm{A} &=& \bra{0}\frac{{\bar q}\gamma^\mu q}{\mathcal{P}^\mu_{\rm V}}\ket{P}_{\rm BT} 
    - \bra{0} \alpha\frac{{\bar q}\sigma^{\mu\nu} q}{\mathcal{P}^{\mu\nu}_{\rm T}}\ket{P}_{\rm BT},\nonumber\\
    f_\rho^\mathrm{S} &=& \bra{0}\frac{{\bar q}\sigma^{\mu\nu} q}{\mathcal{P}^{\mu\nu}_{\rm T}}\ket{P}_{\rm BT} 
    - \bra{0} \alpha\frac{{\bar q}\gamma^\mu q}{\mathcal{P}^\mu_{\rm V}}\ket{P}_{\rm BT}.
\end{eqnarray}
We define
\begin{eqnarray}
\tilde{f}_{\rho}^{\perp} 
&\equiv& \bra{0}\alpha\frac{{\bar q}\sigma^{\mu\nu} q}{\mathcal{P}^{\mu\nu}_{\rm T}}\ket{P}_{\rm BT}\nonumber  \\
    &=& \sqrt{N_c} \int_0^1 {\rm d}x \int \frac{ {\rm d}^2 \bm{k}_\bot}{16\pi^3}\  \Phi(x,\bm{k}_\perp)  \nonumber\\
   & &  \times \mbox{} \frac{2m}{M_0} \frac{1}{\mathcal{P}^{\mu\nu}_{\rm T}}
   \sum_{\lambda_1, \lambda_2} \mathcal{R}_{\lambda_1 \lambda_2}^{11} \left[\frac{\bar{v}_{\lambda_2}(p_2)}{\sqrt{x_2}}
   \sigma^{\mu\nu} \frac{u_{\lambda_1}(p_1)}{\sqrt{x_1}}\right], \quad \quad 
\end{eqnarray}
and 
\begin{eqnarray}
\tilde{f}_{\rho}^{\parallel} 
&\equiv& \bra{0} \alpha\frac{{\bar q}\gamma^{\mu} q}{\mathcal{P}^{\mu}_{\rm V}}\ket{P}_{\rm BT}\nonumber \\
    &=& \sqrt{N_c} \int_0^1 {\rm d}x \int \frac{ {\rm d}^2 \bm{k}_\bot}{16\pi^3}\  \Phi(x,\bm{k}_\perp)  \nonumber\\
   & &  \times \mbox{} \frac{2m}{M_0} \frac{1}{\mathcal{P}^\mu_{\rm V}}
   \sum_{\lambda_1, \lambda_2} \mathcal{R}_{\lambda_1 \lambda_2}^{1h} \left[\frac{\bar{v}_{\lambda_2}(p_2)}{\sqrt{x_2}}
   \gamma^\mu  \frac{u_{\lambda_1}(p_1)}{\sqrt{x_1}}\right].\quad \quad 
\end{eqnarray}
Note that the factor $\alpha=2m/M$ is consistently replaced by $2m/M_0$ within the integrals.
It is important to emphasize that both $\tilde{f}_{\rho}^{\perp}$ and $\tilde{f}_{\rho}^{\parallel}$ are independent of current components
and polarization vectors within our LFQM, validating the self-consistency of our approach.

The final expressions for $\tilde{f}_\rho^\parallel$ and $\tilde{f}_\rho^\perp$ are given by
\be\label{eq:ftilde}
\tilde{f}_{\rho}^{\parallel(\perp)} = \sqrt{N_c}  \int_0^1 {\rm d}x \int \frac{ {\rm d}^2 \bm{k}_\bot}{16\pi^3}\  
	\Phi(x, \bm{k}_\bot) ~\tilde{\mathcal{O}}_{\rm V(T)},
\ee
where $\tilde{\mathcal{O}}_\mathrm{V(T)} = \frac{2m}{M_0}\mathcal{O}_\mathrm{V(T)}$.
We note that in our LFQM, the scale dependence of the decay constants can be approximately incorporated by introducing
an ultraviolet cutoff on the transverse momentum: $\int {\rm d}^2 \bm{k}_\bot\to\int^{|\bm{k}_\bot|<\mu}{\rm d}^2 \bm{k}_\bot$.
For the nonperturbative Gaussian wave function ${\Phi}(x, \bm{k}_\bot)$ given in Eq.~\eqref{HO1S2SJac}, 
we adopt $\mu_0\simeq 1$ GeV as the optimal initial scale in our LFQM.

In our numerical analysis of the $\rho$ meson, 
we use the model parameters $(m, \beta)= (0.25, 0.3194)$ GeV, determined from meson spectrum calculations using 
a HO confining potential and variational analysis within our standard LFQM framework~\cite{CJ99,CJ99B,CJ07}. 
\begin{table}[t]
	\begin{ruledtabular}
		\renewcommand{\arraystretch}{1.5}
		\caption{Predicted decay constants (in MeV) of the $\rho$ meson obtained from various current operators in our LFQM.}
		\label{tab:constant0}
		\begin{tabular}{c|ccc|ccc} 
             $\rho(1S)$   & $f_{\rho}^\parallel$  & $\tilde{f}_\rho^{\perp}$ & $f_{\rho}^\mathrm{A}$ & $f_{\rho}^\perp$ & $\tilde{f}_\rho^{\parallel}$ & $f_{\rho}^\mathrm{S}$ \\ \hline
            $m\neq 0$ & 215 & 93 & 122 & 173 & 112 & 61\\
            $m\to 0$ & 190 & 0 & 190 & 95 & 0 & 95\\ \hline
            Expt.~\cite{ParticleDataGroup:2024cfk} & 208\footnote{Expt. value for $\Gamma(\tau\to\rho\nu_\tau)$.}, 216(5)\footnote{Expt. value for $\rho^0\to e^+e^-$.} & \dots &\dots &\dots  & \dots & \dots \\
      \end{tabular}
		\renewcommand{\arraystretch}{1}
	\end{ruledtabular}
\end{table}

The computed $\rho$ meson decay constants ($f_{\rho}^\parallel$, $f_{\rho}^\perp$, $f_{\rho}^\mathrm{A}$, $f_{\rho}^\mathrm{S}$) and
$\tilde{f}_{\rho}^{\parallel(\perp)}$ are summarized in Table~\ref{tab:constant0}.
The values of $f_{\rho}^\parallel$ and $f_{\rho}^\perp$ have already been analyzed in Refs.~\cite{C13,CJ07,Arifi:2022qnd}.
Our prediction, $f_{\rho}^\parallel=215$ MeV, is in excellent agreement with
the experimental value of $f_{\rho}^{\parallel{(\rm Expt.)}}=216(5)$ MeV, extracted from $\rho^0\to e^+e^-$~\cite{ParticleDataGroup:2024cfk}.

As the main focus of this work, we compute $f_{\rho}^\mathrm{A}$ and $f_{\rho}^\mathrm{S}$. 
These decay constants are related to 
$f_{\rho}^\parallel$ and $f_{\rho}^\perp$ via Eq.~\eqref{eq:fAfS}, which follows from the QCD equation of motion~\cite{Ball:1998sk},
and are expressed in our LFQM as $f_{\rho}^\mathrm{A} =f_{\rho}^\parallel - \tilde{f}_\rho^{\perp}$, 
$f_{\rho}^\mathrm{S} = f_{\rho}^\perp - \tilde{f}_\rho^{\parallel}$, as shown in Eq.~\eqref{eq:fAfSQM}.
These relations are numerically verified in Table~\ref{tab:constant0}.
Furthermore, our model satisfies the SU(6) symmetry relation~\cite{LEUT74},
$f_\pi + f_\rho^\parallel = 2f_\rho^\perp$, as demonstrated in Refs.~\cite{CJ07,Arifi:2023uqc}.
In the chiral limit (i.e., $m\to0$), we also confirm that $f_\rho^\mathrm{A} = f_\rho^\parallel$ and
$f_\rho^\mathrm{S} = f_\rho^\perp$. 
In this limit, the relation $f_\rho^\parallel = 2f_\rho^\perp$ is obtained within our model.

\section{Distribution amplitudes}
\label{sec:das}

In this section, we derive the chiral-even DAs from the nonlocal matrix elements of the current
operators $(\gamma^\mu, \gamma^\mu\gamma_5)$, and the chiral-odd DAs from those of  $( \sigma^{\mu\nu}, \bm{1} )$. 
The twist classification of these DAs depends on the choice of current components and polarization vectors,
as outlined in Ref.~\cite{Ball:1998sk}.
Table~\ref{tab:twist} summarizes all possible combinations of current components and polarization vectors that yield nonvanishing matrix elements.

For example, the chiral-even twist-2 DA, $\phi_{2;\mathrm{V}}^\parallel(x)$, can be obtained from 
the $\gamma^+$ or $\gamma^\perp$ components of the vector current, combined with longitudinal polarization $\epsilon_0$. 
In contrast, using the minus component $\gamma^{-}$ with $\epsilon_0$ leads to the twist-4 DA $\phi_{4;\mathrm{V}}^\parallel$. 
Transverse polarizations $\epsilon_\pm$, combined with $\gamma^-$ or $\gamma^\perp$,
yield the twist-3 DA $\phi_{3;\mathrm{V}}^\perp$. 
All DAs satisfy the normalization condition given in Eq.~\eqref{eq:DAnorm}.
In the following, we present a detailed analysis of both chiral-even and chiral-odd DAs.

\begin{table}[t]
	\begin{ruledtabular}
		\renewcommand{\arraystretch}{2}
		\caption{Twist classification of DAs based on the choice of the chirality, current, component, and polarization for vector mesons.}
		\label{tab:twist}
		\begin{tabular}{ccccc}
		 Chirality &  $\Gamma$    & $\epsilon_h$  & Twist  & DAs \\ \hline 
         Even  &  $\gamma^+,\gamma^\perp$ & $\epsilon_0$  & 2  & $\phi_{2;\mathrm{V}}^\parallel$  \\ 
                     & $\gamma^{-}$  & $\epsilon_0$  &  4     & $\phi_{4;\mathrm{V}}^\parallel$   \\ 
                     & $\gamma^\perp,\gamma^-$ & $\epsilon_\pm$ & 3   & $\phi_{3;\mathrm{V}}^\perp$ \\ 
                     & $\gamma^\perp\gamma_5$ & $\epsilon_\pm$ & 3   & $\psi_{3;\mathrm{A}}^\perp$  \\  \hline 
         Odd  &  $\sigma^{\perp +},\sigma^{+-}$ &  $\epsilon_\pm$ 	& 2  & $\phi_{2;\mathrm{T}}^\perp$    \\ 
                     & $\sigma^{\perp -}$ & $\epsilon_\pm$	& 4  & $\phi_{4;\mathrm{T}}^\perp$    \\ 
                     &$\sigma^{+ -},\sigma^{\perp-}$  & $\epsilon_0$ & 3  & $\phi_{3;\mathrm{T}}^\parallel$\\ 
                     & $\bm{1}$ & $\epsilon_0$  &  3    & $\psi_{3;\mathrm{S}}^\parallel$ \\ 
		\end{tabular}
	\end{ruledtabular}
\end{table}

\subsection{Chiral-even DA}

The chiral-even DAs of a vector meson, associated with the vector current up to twist-4 accuracy, are given by~\cite{Ball:2007zt}
\begin{eqnarray}\label{eq:DA2} A_\mathrm{V}^{\mu}(h)
   &=&\bra{0} \bar{q}(z)\gamma^\mu q(-z)\ket{{\rho}(P,h)} \nonumber\\
   &=& f_{\rho}^\parallel M \int^1_0 \dd x\ {\rm e}^{-i\zeta P\cdot z} \nonumber\\
   & & \times \biggl[ 
   P^\mu \frac{\epsilon_h\cdot z}{P \cdot z} \left( \phi_{2;{\rm V}}^ \parallel(x) + z^2 (\cdots)\right) \nonumber\\
   &&+ \left(\epsilon_h^\mu - P^\mu \frac{\epsilon_h\cdot z}{P \cdot z} \right) \phi_{3;{\rm V}}^\perp(x) -\frac{z^\mu M^2\epsilon_h\cdot z}{2(P \cdot z)^2} \nonumber\\
   & & \times  \left(\phi_{4;{\rm V}}^\parallel(x) +\phi_{2;{\rm V}}^\parallel(x) -2 \phi_{3;{\rm V}}^\perp(x)   \right) \biggr], 
\end{eqnarray}
where the term proportional to $z^2$ vanishes under the equal LF time condition, i.e. for a lightlike separation $z^\mu$ with $z^2 = z^- z^+ - \bm{z}_\perp^2 = 0$, which is realized by setting $z^+ = 0$ and $\bm{z}_\perp = 0$.
Then, Eq.~(\ref{eq:DA2}) reduces to
\begin{eqnarray}
A_\mathrm{V}^{\mu}(h) &=& f_{\rho}^\parallel M \int^1_0 \dd x\ {\rm e}^{-i\zeta P\cdot z} \biggl[P^\mu \frac{\epsilon_h^+ }{P^+ } \phi_{2;{\rm V}}^\parallel(x)  \nonumber\\
& & + \left(\epsilon_h^\mu - P^\mu \frac{\epsilon_h^+ }{P^+} \right) \phi_{3;{\rm V}}^\perp(x) -\frac{z^\mu}{z^-} \frac{M^2\epsilon_h^+ }{P^+P^+ }\nonumber\\
& & \times \left(\phi_{4;{\rm V}}^\parallel(x) +\phi_{2;{\rm V}}^\parallel(x) -2 \phi_{3;{\rm V}}^\perp(x)   \right) \biggr].
\end{eqnarray}

As summarized in Table~\ref{tab:twist}, the twist-2 DA $\phi_{2;{\rm V}}^ \parallel(x)$ can be extracted from 
either the $\mu=+$ or $\mu=\perp$ component of the vector current with longitudinal polarization $(h=0)$:
\be
A_\mathrm{V}^{+(\perp)} (0)
= f_{\rho}^\parallel M \epsilon^{+(\perp)}_{0} \int^1_0 \dd x\ {\rm e}^{-i\zeta P\cdot z} \phi_{2;{\rm V}}^ \parallel(x).
\ee
The twist-3 DA $\phi_{3;{\rm V}}^\perp(x)$ can be isolated from either the $\mu=\perp$ or $\mu=-$ component with transverse polarization $(h=\pm1)$:
\begin{eqnarray}
A_\mathrm{V}^{\perp(-)}(\pm1) &=& f_{\rho}^\parallel M \epsilon^{\perp(-)}_{\pm}  \int^1_0 \dd x\ {\rm e}^{-i\zeta P\cdot z} \phi_{3;{\rm V}}^\perp(x).\quad
\end{eqnarray}
Similarly, the twist-4 DA $\phi_{4;{\rm V}}^\parallel(x)$ can be extracted from the $\mu=-$ component with longitudinal polarization $(h=0)$ in the $\bm{P}_\perp=0$ frame:
\begin{eqnarray}
A_\mathrm{V}^-(0) 
&=& f_{\rho}^\parallel M \epsilon^-_{0}\int^1_0 \dd x\ {\rm e}^{-i\zeta P\cdot z}  \phi_{4;{\rm V}}^\parallel(x).  \quad \quad 
\end{eqnarray}

The explicit forms of the three chiral-even DAs,
$\phi_\mathrm{V} = \left\{\phi_{2;{\rm V}}^\parallel, \phi_{3;{\rm V}}^\perp, \phi_{4;{\rm V}}^\parallel \right\}$,
derived from the vector current, are given by
\begin{eqnarray}
    \phi_\mathrm{V}(x) &=& \frac{\sqrt{N_c} }{f_\rho^\parallel}  \int \frac{\dd^2 \bm{k}_\bot}{16\pi^3}\ \Phi(x,\bm{k}_\perp) \mathcal{O}_{\rm V},
\end{eqnarray}
where the corresponding operators $\mathcal{O}_{\rm V}$, defined for each DA, are
\begin{eqnarray}
\mathcal{O}_\mathrm{V} &=& 
\left\{\mathcal{O}_{\rm V}^{+(\perp)}(0), \mathcal{O}_{\rm V}^{\perp(-)}(\pm 1), \mathcal{O}_{\rm V}^{-}(0) \right\}.
\end{eqnarray}
Their explicit forms are listed in Table~\ref{tab:t1}.
Notably, as shown in Table~\ref{tab:t1}, for the $\rho$ meson, the three operators $\mathcal{O}_{\rm V}^+(0)$, $\mathcal{O}_{\rm V}^{\perp}(0)$, 
and $ \mathcal{O}_{\rm V}^{-}(0)$
coincide. As a result, the twist-2 DA $\phi_{2;{\rm V}}^\parallel$ and the twist-4 DA $\phi_{4;{\rm V}}^\parallel$ are identical in our model. 

Finally, the chiral-even twist-3 DA $\psi_{3;\mathrm{A}}^\perp(x)$, derived from the nonlocal axial-vector matrix element  
in Eq.~\eqref{eq:nonlocal1}, is expressed in Eq.~\eqref{eq:psi3_a} within our LFQM. 
Applying the BT construction (i.e. $M\to M_0$) yields the final form:
\begin{eqnarray}
\psi_{3;\mathrm{A}}^\perp(x) &=& \frac{\sqrt{N_c}}{f_\rho^\mathrm{A}}\int_0^x \dd x^\prime  
\int \frac{{\rm d}^2\bm{k}_\perp }{16\pi^3} \Phi(x^\prime,\bm{k}_\perp) \mathcal{O}_\mathrm{A}^{\prime}, \quad 
\end{eqnarray}
where $\mathcal{O}_\mathrm{A}^{\prime}=2\mathcal{O}_\mathrm{S}^{\prime}$ (see Eq.~\eqref{eq:Os}).


\subsection{Chiral-odd DA}

The chiral-odd DAs of a vector meson, associated with the tensor current up to twist-4 accuracy, are given by~\cite{Ball:2007zt}
\begin{eqnarray}\label{eq:DAodd}
   A_\mathrm{T}^{\mu\nu}(h) &=&\bra{0} \bar{q}(z)\sigma^{\mu\nu} q(-z)\ket{\rho(P,h)} \nonumber\\
   &=& if_\rho^\perp \int^1_0 \dd x\ {\rm e}^{-i\zeta P\cdot z} \biggl\{ \left( \epsilon^\mu_h P^\nu - \epsilon^\nu_h P^\mu \right)\nonumber\\
   &&\times \left[ \phi_{2;\mathrm{T}}^ \perp(x) + z^2 (\cdots ) \right] + \left(P^\mu z^\nu - P^\nu z^\mu\right)  \nonumber\\
   && \times \frac{(\epsilon_h \cdot z)M^2}{(P\cdot z)^2}  \left[\phi_{3;\mathrm{T}}^\parallel(x) -\tfrac{1}{2}\phi_{2;\mathrm{T}}^ \perp(x) -\tfrac{1}{2}\phi_{4;\mathrm{T}}^ \perp(x) \right]\nonumber\\
   &&+ \frac{\left( \epsilon^\mu_h z^\nu - \epsilon^\nu_h z^\mu \right)  M^2}{P\cdot z}  \biggl[\phi_{4;\mathrm{T}}^\perp(x) -\phi_{2;\mathrm{T}}^\perp(x)  \biggr] \biggr\}. \nonumber\\
\end{eqnarray}
As in Eq.~\eqref{eq:DA2}, the term proportional to $z^2$ vanishes under the equal LF time condition, i.e., when $z^+ = \bm{z}_\perp = 0$. This simplifies the expression to
\begin{eqnarray}
A_\mathrm{T}^{\mu\nu}(h) &=& if_\rho^\perp \int^1_0 \dd x\ {\rm e}^{-i\zeta P\cdot z}  \biggl\{ \left( \epsilon^\mu_h P^\nu - \epsilon^\nu_h P^\mu \right) \phi_{2;\mathrm{T}}^ \perp(x)\nonumber\\
   && +  \frac{2M^2\epsilon_h^+\left(P^\mu z^\nu - P^\nu z^\mu\right)}{z^-P^+P^+} \nonumber\\
   &&\times \left[\phi_{3;\mathrm{T}}^\parallel(x) -\tfrac{1}{2}\phi_{2;\mathrm{T}}^ \perp(x) -\tfrac{1}{2}\phi_{4;\mathrm{T}}^ \perp(x) \right]\nonumber\\
   &&+ \frac{2 M^2 \left( \epsilon^\mu_h z^\nu - \epsilon^\nu_h z^\mu \right)}{P^+z^-}  \biggl[\phi_{4;\mathrm{T}}^\perp(x) -\phi_{2;\mathrm{T}}^\perp(x)  \biggr] \biggr\}. \nonumber\\ 
\end{eqnarray}
As summarized in Table~\ref{tab:twist}, the twist-2 DA $\phi_{2;\mathrm{T}}^\perp(x)$ can be isolated by choosing $\mu\nu=\perp +$ or $+-$ with transverse polarization ($h=\pm1$), resulting in
\be
A_\mathrm{T}^{\perp+} (\pm1)
= if_\rho^\perp \epsilon^\perp_\pm P^+ \int^1_0 \dd x\ {\rm e}^{-i\zeta P\cdot z} \phi_{2;\mathrm{T}}^ \perp(x).  
\ee
Similarly, the twist-3 DA $\phi_{3;{\rm T}}^ \perp(x)$ can be extracted by taking $\mu\nu=+-$ or $\perp-$ with longitudinal polarization ($h=0$), yielding
\be
A_\mathrm{T}^{+-} (0)
= if_\rho^\perp \epsilon^+_0 \frac{2M^2}{P^+} \int^1_0 \dd x\ {\rm e}^{-i\zeta P\cdot z} \phi_{3;{\rm T}}^ \perp(x).  
\ee
The twist-4 DA $\phi_{4;\mathrm{T}}^\perp(x)$ is obtained by choosing $\mu\nu=\perp-$ with transverse polarization ($h=\pm1$), leading to
\be
A_\mathrm{T}^{\perp-} (\pm1)
= if_\rho^\perp \epsilon^\perp_\pm \frac{2M^2}{P^+} \int^1_0 \dd x\ {\rm e}^{-i\zeta P\cdot z} \phi_{4;\mathrm{T}}^\perp(x).  
\ee
The three chiral-odd DAs, $\phi_\mathrm{T} = \left\{\phi_{2;{\rm T}}^\perp, \phi_{3;{\rm T}}^\parallel, \phi_{4;{\rm T}}^\perp \right\}$,
derived from the tensor current, can be expressed as
\begin{eqnarray}
    \phi_\mathrm{T}(x) &=& \frac{\sqrt{N_c} }{f_\rho^\perp}  \int \frac{\dd^2 \bm{k}_\bot}{16\pi^3}\ 
    \Phi(x,\bm{k}_\perp)  \mathcal{O}_{\rm T}, 
\end{eqnarray}
where the corresponding operators $\mathcal{O}_{\rm T}$, defined for each DA, are
\begin{eqnarray}
\mathcal{O}_\mathrm{T} &=& \left\{\mathcal{O}_{\rm T}^{\perp+}(\pm1), \mathcal{O}_{\rm T}^{+-}(0), \mathcal{O}_{\rm T}^{\perp-}(\pm1) \right\}.
\end{eqnarray}
The explicit forms of these operators are provided in Table~\ref{tab:t1}. 
Similar to the chrial-even case,  the two operators $\mathcal{O}_{\rm T}^{\perp+}(\pm1)$ and $\mathcal{O}_{\rm T}^{\perp-}(\pm1)$
coincide. As a result, the twist-2 DA $\phi_{2;{\rm T}}^\perp$ and the twist-4 DA $\phi_{4;{\rm T}}^\perp$ are identical in our model.

Finally, the chiral-odd twist-3 DA $\psi_{3;\mathrm{S}}^\parallel(x)$, derived from the nonlocal scalar matrix element  
in Eq.~\eqref{eq:nonlocal2}, is expressed in Eq.~\eqref{eq:psi3_s} within our LFQM. 
Applying the BT construction (i.e. $M\to M_0$) yields the final form:
\begin{eqnarray}
\psi_{3;\mathrm{S}}^\parallel(x) &=& \frac{\sqrt{N_c}}{f_\rho^\mathrm{S}}\int_0^x \dd x^\prime \int \frac{{\rm d}^2\bm{k}_\perp }{2(2\pi)^3} 
\Phi(x^\prime,\bm{k}_\perp)  \mathcal{O}^\prime_\mathrm{S}.
\quad 
\end{eqnarray}


\subsection{Numerical Results}

Parts of results for the chiral-even twist-2 and twist-3 DAs of the $\rho$ meson 
were previously presented in Ref.~\cite{C13}. 
In this study, we complete the analysis by including DAs up to twist-4 for both chiral-even and chiral-odd sectors. 

\begin{figure}[t]
	\centering
	\includegraphics[width=1\columnwidth]{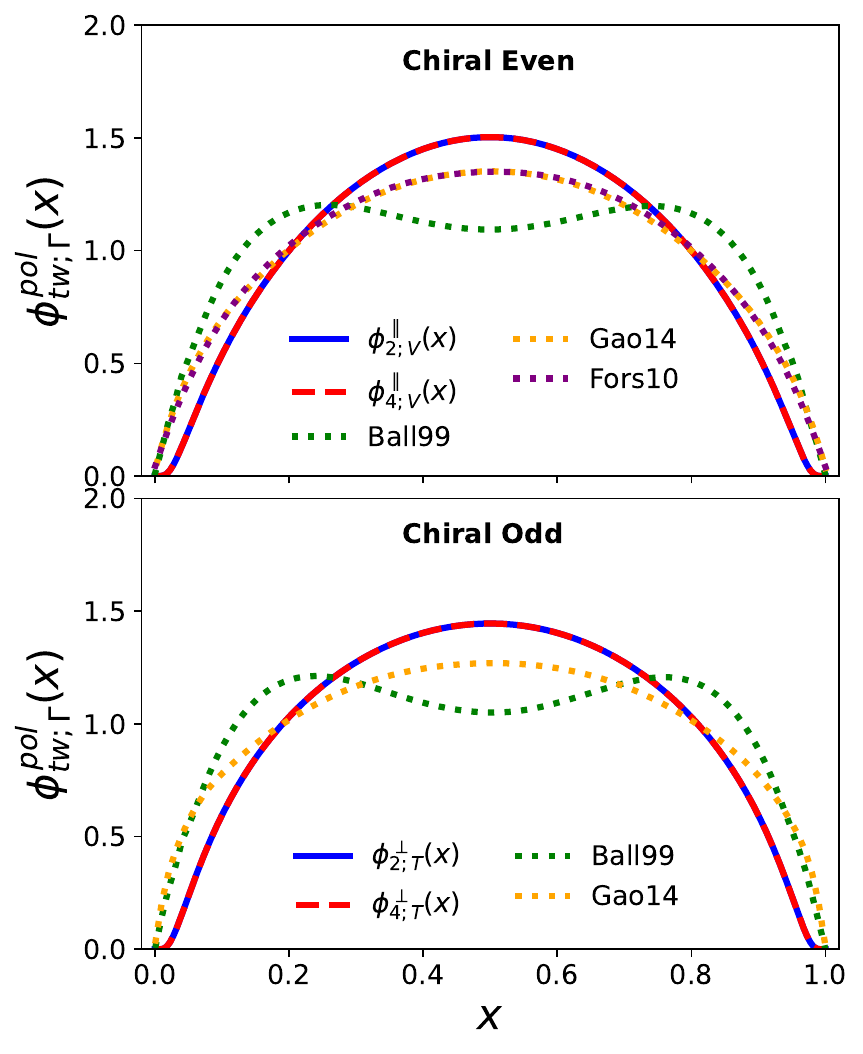}
	\caption{\label{fig:da_tw2} Chiral-even (upper panel) and chiral-odd (lower panel) twist-2 and twist-4 DAs of the $\rho$ meson from our LFQM, compared with twist-2 DAs from other models, including QCD SRs~\cite{Ball:1998ff}, DSE approach~\cite{Gao:2014bca}, and HERA-fit\cite{Forshaw:2010py}. }
\end{figure}

Figure~\ref{fig:da_tw2} shows the chiral-even twist-2 and twist-4 DAs ($\phi_{2;\mathrm{V}}^\parallel, \phi_{4;\mathrm{V}}^\parallel$) (upper panel), and the chiral-odd 
twist-2 and twist-4 DAs ($\phi_{2;\mathrm{T}}^\perp, \phi_{4;\mathrm{T}}^\perp$) (lower panel) for the $\rho$ meson. The twist-2 results are
also compared with other model calculations~\cite{Ball:1998ff,Gao:2014bca,Forshaw:2010py}.

We note that in our LFQM, the twist-2 and twist-4 DAs coincide and follow the asymptotic form $\phi_\mathrm{asm}(x) = 6x(1-x)$
in the case of equal quark masses. However, the agreement in this case is fortuitous as the twist-2 and twist-4 DAs indeed differ for mesons with unequal quark masses, 
reflecting the sensitivity of higher-twist structures to mass asymmetry even within the valence $q{\bar q}$ framework.
For the twist-2 DA, the single-humped structure we obtain is consistent with---but slightly narrower than--- that 
from the DSE approach~\cite{Gao:2014bca} and the HERA-fit~\cite{Forshaw:2010py}, while it differs significantly from the double-humped 
structure predicted by QCD SRs~\cite{Ball:1998ff}.
Our DA is more sharply peaked around $x=1/2$ and more strongly suppressed near the endpoints 
$x=0$ and $x=1$, which may be attributed to the use of a Gaussian radial wave function in our model that naturally favors the central momentum region.

\begin{figure}[t]
	\centering
	\includegraphics[width=1\columnwidth]{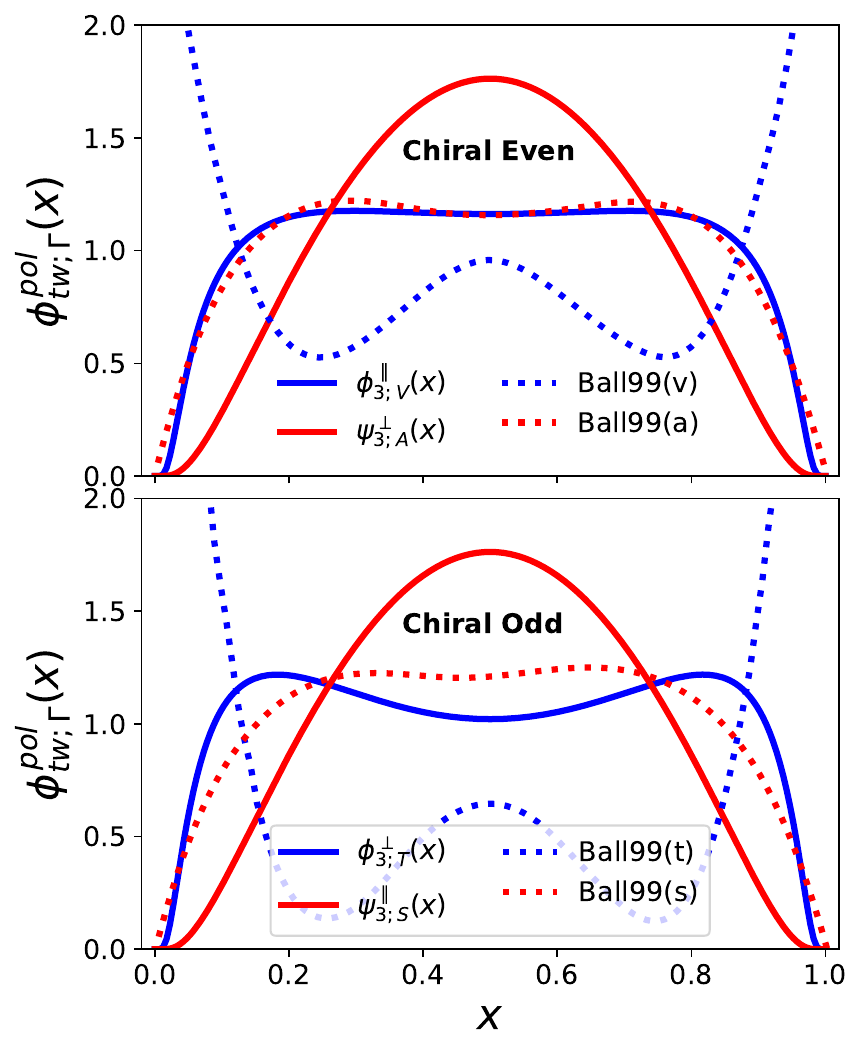}
	\caption{\label{fig:da_tw3} The chiral-even (upper panel) and chiral-odd (lower panel) twist-3 DAs of the $\rho$ meson from our LFQM,
    compared with those from QCD SRs~\cite{Ball:1998ff}.}
\end{figure}

\begin{figure}[t]
	\centering
	\includegraphics[width=1\columnwidth]{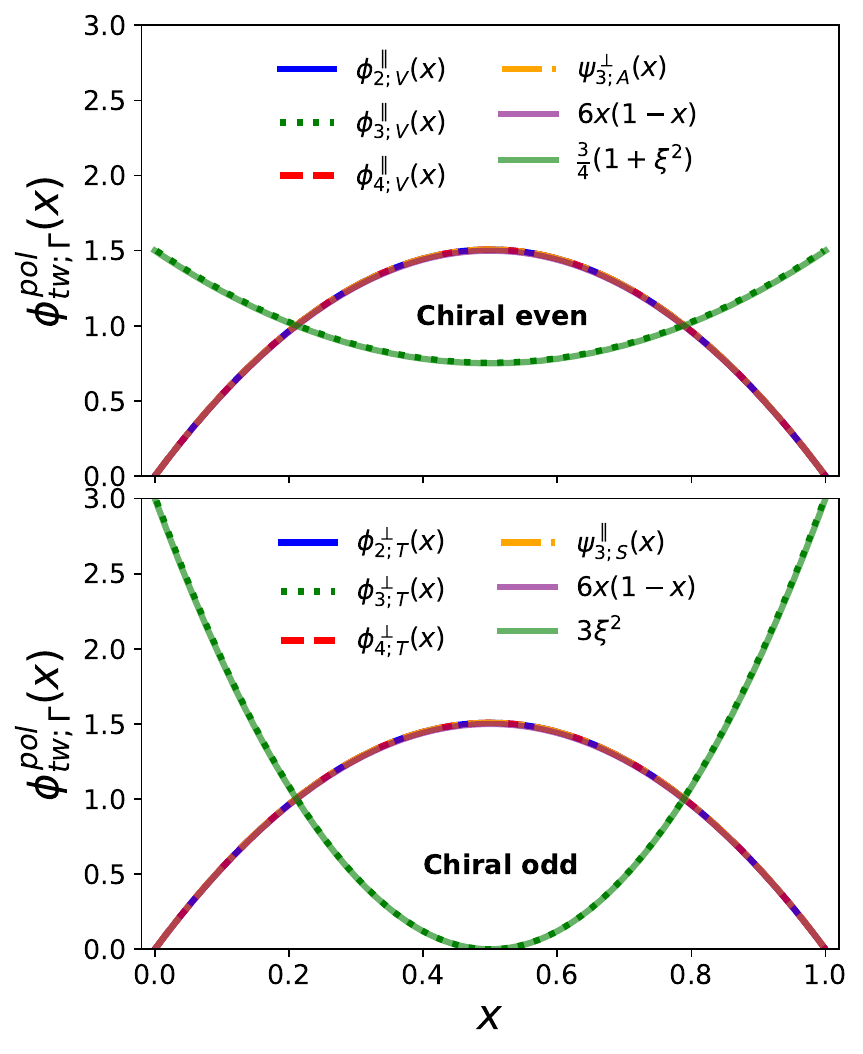}
	\caption{\label{fig:da_chiral} Chiral-even (upper panel) and chiral-odd (lower panel) DAs of the $\rho$ meson 
    in the chiral limit $(m\to 0)$, calculated in our LFQM. The twist-2 and twist-4, $\psi_{3;A}^\perp(x)$, along with $\psi_{3;S}^\parallel(x)$ DAs match the asymptotic form $6x(1 - x)$. The twist-3 DAs, $\phi_{3;V}^\parallel(x)$ and $\phi_{3;T}^\perp(x)$,
    match the asymptotic forms $\frac{3}{4}(1 + \xi^2)$ and $3\xi^2$, respectively, as predicted by QCD SRs~\cite{Ball:1998ff}. 
    }
\end{figure}

Figure~\ref{fig:da_tw3} presents the twist-3 DAs of the $\rho$ meson.
Our results are also compared with those obtained from QCD SRs~\cite{Ball:1998ff}.
The chiral-even components (upper panel), $\phi_{3;\mathrm{V}}^\parallel(x)$ and $\psi_{3;\mathrm{A}}^\perp(x)$, exhibit distinct shapes: 
$\phi_{3;\mathrm{V}}^\parallel(x)$ shows a broad double-humped profile, while $\psi_{3;\mathrm{A}}^\perp(x)$ displays a narrow, single-peaked distribution 
centered at $x=1/2$. Compared to the QCD sum rule predictions~\cite{Ball:1998ff}, our $\psi_{3;\mathrm{A}}^\perp(x)$ is notably more localized, 
and $\phi_{3;\mathrm{V}}^\parallel(x)$ differs substantially, lacking the end-point enhancement and double-minimum structure seen in~\cite{Ball:1998ff}. 
Interestingly, the shape of our $\phi_{3;\mathrm{V}}^\parallel(x)$ more closely resembles that of $\psi_{3;\mathrm{A}}^\perp(x)$ from QCD SRs, though 
their asymptotic behaviors remain distinct (see Fig.~\ref{fig:da_chiral}).

The chiral-odd twist-3 DAs (lower panel), $\phi_{3;\mathrm{T}}^\perp(x)$ and $\psi_{3;\mathrm{S}}^\parallel(x)$, follow similar patterns: 
$\phi_{3;\mathrm{T}}^\perp(x)$ shows a more pronounced double-humped structure with deeper suppression at the center, while $\psi_{3;\mathrm{S}}^\parallel(x)$ 
has a single central peak, akin to $\psi_{3;\mathrm{A}}^\perp(x)$. 
These results again contrast with the QCD SRs~\cite{Ball:1998ff}, highlighting the model dependence of twist-3 DAs.

In Fig.~\ref{fig:da_chiral}, we present the chiral-even (upper panel) and chiral-odd (lower panel) DAs of the $\rho$ meson up to twist-4 in the chiral limit $(m\to 0)$.
It is evident that our chiral-limit predictions for the chiral-even DAs ($\phi_{2,\mathrm{V}}^\parallel(x)$, $\phi_{4,\mathrm{V}}^\parallel(x)$, and $\psi_{3,\mathrm{A}}^\perp(x)$) coincide and match
the asymptotic form $6x(1-x)$. The same holds for the chiral-odd DAs ($\phi_{2,\mathrm{T}}^\perp(x)$, $\phi_{4,\mathrm{T}}^\perp(x)$, and $\psi_{3,\mathrm{S}}^\parallel(x)$) which also coincide 
and match the asymptotic shape $6x(1-x)$.
In contrast, the twist-3 DAs
$\phi_{3,\mathrm{V}}^\parallel(x)$ and $\phi_{3,\mathrm{T}}^\perp(x)$ exhibit 
characteristic convex shapes in the chiral limit, consistent with the corresponding asymptotic predictions from QCD SRs~\cite{Ball:1998sk}.
Specifically, $\phi_{3,\mathrm{T}}^\perp(x)$ matches the asymptotic form $3\xi^2$, where $\xi=2x-1$,
attaining a value of 3 at the endpoints ($x=0,1$) and vanishing at $x=1/2$, 
while $\phi_{3,\mathrm{V}}^\parallel(x)$ follows the asymptotic form $\frac{3}{4}(1+\xi^2)$, reaching 3/2 at the endpoints and 3/4 at $x=1/2$. 
These results demonstrate that our LFQM predictions for twist-3 DAs in the chiral limit accurately reproduce the expected asymptotic behavior.

\begin{table*}[t]
    \renewcommand{\arraystretch}{1.2}
    \caption{The $\xi$ and Gegenbaur moments up to $n=6$ for chiral-even and -odd DAs of the $\rho$ meson up to twist 4 compared with other model calculations. The odd moments are vanishing because of the equal mass of the constituent quark.}
    \label{tab:moment}
    \begin{tabular}{l|l|ccc|l|ccc}
        \hline\hline
        DA & & $\expval{\xi^2}$ & $\expval{\xi^4}$ & $\expval{\xi^6}$&  & $a_2$ & $a_4$ & $a_6$ \\ \hline
        $\phi_{2;\mathrm{V}}^\parallel$ 
            & Our & 0.193 & 0.078 & 0.040 & Our & $-0.020$ & $-0.030$ & $-0.017$ \\
            & HERA-fit~\cite{Forshaw:2010py}  & 0.227 & 0.105 & 0.062 & QCD SRs~\cite{Ball:1998sk} & 0.18(10) & \dots & \dots\\
            & QCD SRs~\cite{Zhong:2023cyc} & 0.220(6) & 0.103(4) & 0.0656(50) & QCD SRs~\cite{Pimikov:2013usa} &  0.047(58) & $-0.057(118)$ & \dots \\
            & DSE~\cite{Gao:2014bca} & 0.23 & 0.11 & 0.066 &  QCD SRs~\cite{Ball:2004rg} & $0.09^{+0.10}
_{-0.07}$ & $0.03(2)$ & \dots \\
            & LFH~\cite{Gurjar:2024wpq} & 0.20 & 0.087 & 0.048 & LFQM~\cite{Ji:1992yf} & $-0.03$ & $-0.09$ &0.7 \\
            & Lattice QCD~\cite{Arthur:2010xf} & 0.25(2)(2) & \dots & \dots & Lattice QCD~\cite{Braun:2016wnx} & 0.132(27) & \dots & \dots \\
        $\phi_{3;\mathrm{V}}^\parallel$ 
            & Our & 0.254 & 0.120 & 0.069 & Our & 0.158 & $-0.018$ & $-0.036$ \\
        $\psi_{3;\mathrm{A}}^\perp$ 
            & Our & 0.153 & 0.052 & 0.023 & Our & $-0.138$ & $-0.023$ & $-0.002$ \\
        $\phi_{4;\mathrm{V}}^\parallel$ 
            & Our & 0.193 & 0.078 & 0.040 & Our & $-0.020$ & $-0.030$ & $-0.017$ \\ \hline 
        $\phi_{2;\mathrm{T}}^\perp$ 
            & Our & 0.202 & 0.084 & 0.044 & Our & 0.007 & $-0.032$ & $-0.020$ \\
            & LFH~\cite{Gurjar:2024wpq} & 0.25 & 0.13 & 0.079 & QCD SRs~\cite{Ball:1998sk} & 0.2(1) & \dots & \dots \\
            & DSE~\cite{Gao:2014bca} & 0.25 & 0.13 & 0.079 &  QCD SRs~\cite{Ball:2004rg} & $0.09^{+0.10}
_{-0.07}$ & $0.03(2)$ & \dots \\
 & QCD SRs~\cite{Bakulev:1998pf} & 0.325(10) & \dots & \dots & LFQM~\cite{Ji:1992yf} & 0 & $-0.04$ & $-0.04$ \\
 & QCD SRs~\cite{Pimikov:2013usa} & 0.11(1) & 0.022(2) & \dots & Lattice QCD~\cite{Braun:2016wnx} & 0.101(22) & \dots & \dots \\
        $\phi_{3;\mathrm{T}}^\perp$ 
            & Our & 0.278 & 0.136 & 0.080 & Our & 0.228 & 0.017 & $-0.044$ \\
        $\psi_{3;\mathrm{S}}^\parallel$ 
            & Our & 0.153 & 0.084 & 0.023 & Our & $-0.138$ & $-0.023$ & $-0.002$ \\
        $\phi_{4;\mathrm{T}}^\perp$ 
            & Our & 0.202 & 0.052 & 0.044 & Our & 0.007 & $-0.032$ & $-0.020$ \\ 
        \hline\hline
    \end{tabular}
    \renewcommand{\arraystretch}{1}
\end{table*}

As seen in Fig.~\ref{fig:da_chiral}, the leading- and higher-twist DAs in the chiral limit ($m\to 0$) approach two types of asymptotic 
forms—namely, $6x (1-x)$ and $\frac{3}{4}(1+\xi^2)$—depending  on the twist order.
In contrast, in the opposite limit of the heavy-quark regime ($m \to \infty$), all DAs converge to a common form proportional to $\delta(x-1/2)$,
independent of twist.  This behavior arises from the fact that the LFWF in the heavy-quark limit becomes
\begin{equation}
\phi(x,\bm{k}_\perp) \propto \delta\left(x - \tfrac{1}{2}\right) e^{-\frac{\bm{k}_\perp^2}{2\beta^2}},
\end{equation}
indicating a sharp localization of the longitudinal momentum fraction at $x = 1/2$.  
As a result, the twist-2, -3, and -4 DAs all collapse into a single delta-function-like profile centered at $x = 1/2$,
reflecting the suppression of relativistic effects in the heavy-quark limit and the non-relativistic reduction of the internal meson dynamics.
Such universal behavior provides a useful theoretical benchmark and differs clearly from the richer structures observed 
in the chiral and intermediate mass regimes. 
Further analysis of heavy quarkonia DAs in connection with the non-relativistic limit can be found in Ref.~\cite{PQCDCJ}.

To facilitate comparison, we compute the $\xi$-moments up to order $n=6$, defined as
\begin{equation}
    \expval{\xi^n} = \int_0^1 {\rm d}x\ \xi^n\ \phi(x),
\end{equation}
where $\phi(x)$ represents the DA of the $\rho$ meson.
These DAs can also be expanded in terms of Gegenbauer polynomials $C_n^{3/2}$ as
\begin{equation}
    \phi(x,\mu)=6x(1-x)\left[1 + \sum_{n=1}^\infty a_n(\mu) C_n^{3/2}(\xi) \right],
\end{equation}
with the coefficients $a_n(\mu)$ are known as Gegenbauer moments.
The $\xi$-moments and Gegenbauer moments are related through specific integral relations~\cite{C15}.

Table~\ref{tab:moment} presents the calculated $\xi$-moments and the corresponding Gegenbauer moments $a_n(\mu)$ at the scale $\mu=1$ GeV, 
along with results from other theoretical approaches for comparison.
Our results for the leading-twist DAs show good agreement with those from the light-front holographic (LFH) model~\cite{Gurjar:2024wpq}, QCD SRs~\cite{Zhong:2023cyc}, and DSE approach~\cite{Gao:2014bca}, 
but differ from the lattice QCD results~\cite{Arthur:2010xf,Braun:2016wnx}. 

Of special interest, using the soft pion theorem, crossing symmetry, and dispersion relations for the two-pion distribution amplitude ($2\pi$DA), the authors in Ref.~\cite{Polyakov:2020cnc} argued that the second Gegenbauer moment  
for the leading-twist chiral-even DA $\phi_{2;\mathrm{V}}^\parallel$ is most likely negative. 
Their phenomenological analysis predicts the ratio 
$a_2^\rho / a_2^\pi = (-1.15 \pm 0.86)(1.0\pm 0.1)\in[-2.211,-0.261]$. Our model prediction yields $a_2^\rho = -0.020$ and 
$a_2^\pi=0.050$~\cite{CJ07}, resulting in $a_2^\rho / a_2^\pi=-0.400$, thereby satisfying this constraint.
While some other models predict a positive sign for $a_2$, this discrepancy 
may stem from the proximity of the DA to its asymptotic form and the strong 
sensitivity to the choice of model LFWFs.
Finally, we note that although leading-twist DAs have been extensively studied, 
reliable predictions for higher-twist DAs remain limited.

\section{Conclusion} 
\label{sec:summary}

In this study, we have presented a comprehensive analysis of the $\rho$ meson decay constants and chiral-even and chiral-odd DAs up to twist 4, within the framework of the standard LFQM based on the BT construction~\cite{BT53,KP91}. 
The BT construction provides a consistent framework by incorporating interactions into the Casimir mass operator, ensuring compliance with 
the Poincar\'{e} group symmetry, and allowing the description of mesons as bound states of non-interacting quarks and antiquarks. 

We note that, within the BT construction adopted in our LFQM, the total LF momentum operator is designed to satisfy 
the Poincar\'e algebra, ensuring Lorentz symmetry at the level of the wave function. 
Consequently, quantities such as decay constants and DAs extracted from different current components or polarization states are expected to agree by construction. 
Our explicit calculations thus serve as a numerical confirmation of this self-consistency, verifying that our implementation faithfully preserves the anticipated symmetry properties.

For the $\rho$ meson, which exhibits both longitudinal and transverse polarizations,
we have computed all relevant decay constants $(f_\rho^\parallel,f_\rho^\perp,f_\rho^\mathrm{A},f_\rho^\mathrm{S})$
associated with various current operators $\Gamma=(\gamma^\mu, \sigma^{\mu\nu},\gamma^\mu\gamma_5,\bm{1})$. As these decay constants are physical observables, their theoretical predictions must be entirely independent of the computational convenience of choosing current components, polarization states and reference frames.    
For the first time, we have demonstrated this self-consistency within the LFQM in the computation of these decay constants.

The extraction of decay constants from axial-vector and scalar currents posed a particularly nontrivial challenge due to the mixing between 
$f_\rho^\parallel$ and $f_\rho^\perp$ arising from quark mass corrections, expressed as $f_\rho^\mathrm{A} = f_\rho^\parallel - \alpha f_\rho^\perp$ 
and $f_\rho^\mathrm{S} = f_\rho^\perp - \alpha f_\rho^\parallel$, with $\alpha = 2m/M$. 
By applying the BT construction consistently, 
we resolved these mixings without encountering singularities, ensuring self-consistent robust and reliable results. 
This work complements and extends our earlier studies~\cite{Arifi:2022qnd,Arifi:2023uqc}, further confirming
the applicability of the BT construction beyond two-point functions and 
demonstrating its utility in three-point and higher-order processes~\cite{CJ2024:pion,Ridwan:2024ngc}.

Furthermore, we have predicted the full set of chiral-even DAs ($\phi_{2;\mathrm{V}}^\parallel, \phi_{3;\mathrm{V}}^\parallel, \psi_{3;\mathrm{A}}^\perp, \phi_{4;\mathrm{V}}^\parallel$) 
and chiral-odd DAs ($\phi_{2;\mathrm{T}}^\perp, \phi_{3;\mathrm{T}}^\perp, \psi_{3;\mathrm{S}}^\parallel, \phi_{4;\mathrm{T}}^\perp$) beyond the leading twist. 
These eight DAs reveal rich structural information about the $\rho$ meson. 
Notably, in the chiral limit ($m \to 0$), our predicted DAs are consistent with those from QCD SRs~\cite{Ball:1998sk}, lending further credibility to our LFQM.

Despite the complexity of the $\rho$ meson, the methodology and results demonstrated here can be readily extended to other mesons and observables. Future lattice QCD simulations may serve as valuable benchmarks for validating these predictions.

In conclusion, this work presents a self-consistent and process-independent implementation of the LFQM based on the BT framework.
 It offers a significant step forward to providing a reliable tool for the study of hadron structure and spectroscopy.

\section*{Acknowledgment}
A. J. A acknowledges the hospitality of Kyungpook National University during his stay, which facilitated the completion of this work. 
We thank Kazuhiro Tanaka for useful communications. 
The work of H.-M. C. was supported by the National Research Foundation of Korea(NRF) under Grant No. RS-2023-NR076506.
The work of C.-R.J. was supported in part by the U.S. Department of Energy (Grant No. DE-FG02-03ER41260). 
The National Energy Research Scientific Computing Center (NERSC) supported by the Office of Science of the U.S. Department of Energy 
under Contract No. DE-AC02-05CH11231 is also acknowledged.

\bibliography{references}

\end{document}